\documentclass[10pt]{article}

\usepackage{amsmath}
\usepackage{amssymb}

\usepackage{graphicx}

\usepackage{cite}

\usepackage{color} 

\usepackage{epstopdf}

\usepackage{hyperref}
\hypersetup{pdfborder={0 0 0}} 

\usepackage{fixltx2e}

\usepackage{xcolor}

\usepackage[labelfont=bf,labelsep=period,justification=raggedright]{caption}

\makeatletter
\renewcommand{\@biblabel}[1]{\quad#1.}
\makeatother

\date{}

\usepackage{lastpage,fancyhdr}

\begin{document}
	
	\begin{flushleft}
		{\Large
			
			{Extending integrate-and-fire 
				model neurons to account for the effects of weak electric fields and input filtering mediated by the dendrite}
		}
		\begin{center}
			Florian Aspart$^{1,2,\ast,\ddag}$, 
			Josef Ladenbauer$^{1,2,\ast,\ddag}$,
			Klaus Obermayer$^{1,2}$
		\end{center}
		
		\bf{1} Department of Software Engineering and Theoretical Computer Science, Technische Universit\"at Berlin, Germany
		\\
		\bf{2} Bernstein Center for Computational Neuroscience Berlin, Germany
		
		\vspace{0.25cm}
		$\ast$ These authors contributed equally to this work \\
		$\ddag$ E-mail: florian.aspart@tu-berlin.de, josef.ladenbauer@tu-berlin.de 
	\end{flushleft}
%

%
\section*{Abstract}
Transcranial brain stimulation and evidence of ephaptic coupling have recently sparked strong interests in understanding the effects of weak electric fields on the dynamics of brain networks and of coupled populations of neurons. The collective dynamics of large neuronal populations can be efficiently studied using single-compartment (point) model neurons of the integrate-and-fire (IF) type as their elements.
These models, however, lack the dendritic morphology required to biophysically describe the effect of an extracellular electric field on the neuronal membrane voltage. 
Here, we extend the IF point neuron models to accurately reflect morphology dependent electric field effects extracted from a canonical spatial ``ball-and-stick'' (BS) neuron model.
Even in the absence of an extracellular field, neuronal morphology by itself strongly affects the cellular response properties. We, therefore, derive additional components for leaky and nonlinear IF neuron models to reproduce the subthreshold voltage and spiking dynamics of the BS model exposed to both fluctuating somatic and dendritic inputs and an extracellular electric field.  
We show that an oscillatory electric field causes spike rate resonance, or equivalently, pronounced spike to field coherence. Its resonance frequency depends on the location of the synaptic background inputs. For somatic inputs the resonance appears in the beta and gamma frequency range, whereas for distal dendritic inputs it is shifted to even higher frequencies. 
Irrespective of an external electric field, the presence of a dendritic cable attenuates the subthreshold response at the soma to slowly-varying somatic inputs while implementing a low-pass filter for distal dendritic inputs.
Our point neuron model extension is straightforward to implement and is computationally much more efficient compared to the original BS model. It is well suited for studying the dynamics of large populations of neurons with heterogeneous dendritic morphology with (and without) the influence of weak external electric fields.
%

%
%
%

%
\section*{Author Summary}
How extracellular electric fields -- as generated endogenously or through transcranial brain stimulation -- affect the dynamics of neuronal populations is of great interest but not well understood. 
To study neuronal activity at the network level single-compartment neuron models have been proven very successful, because of their computational efficiency and analytical tractability.
Unfortunately, these models lack the dendritic morphology to biophysically account for the effects of electric fields, and for changes in synaptic integration due to morphology alone.
Here, we consider a canonical, spatially extended %
model neuron and characterize its responses to fluctuating synaptic input as well as an oscillatory, weak electric field. 
In order to accurately reproduce these responses we analytically derive an extension for the popular integrate-and-fire point neuron models.
We show that the dendritic cable acts as a filter for the synaptic input current, which depends on the input location, and that an electric field modulates the neuronal spike rate strongest at a certain (preferred) field frequency.
These phenomena can be successfully reproduced using integrate-and-fire models, extended by a small number of components that are straightforward to implement.  
The extended point models are thus well suited for studying populations of coupled neurons with different morphology, exposed to extracellular electric fields.

%

%
%
%
%

%

%
%
%
%
%
%
%
%

%
%
%

%
%
%
%
%
%
%
%

\thispagestyle{empty}
\clearpage
\setcounter{page}{1}
\pagestyle{fancy}
\fancyhf{}
\lhead{IF neuron model extension: electric fields and dendritic input filter}
\rhead{Aspart, Ladenbauer, Obermayer}
\rfoot{\thepage/\pageref{LastPage}}

%

%
\section*{Introduction}

Extracellular electric fields in the brain and their impact on neural activity have gained a considerable amount of attention in neuroscience over the past decade. 
These electric fields can be generated endogenously \cite{Frohlich2010,Buzsaki2012,Einevoll2013} or through transcranial (alternating) current stimulation \cite{Neuling2012,Datta2009,Bikson2012a}, and can modify the activity of neuronal populations in various ways \cite{Frohlich2010,Reato2010,Anastassiou2011,Ali2013}. Although the fields generated by this type of noninvasive brain stimulation are rather weak ($\leq 1~\mathrm{V}/\mathrm{m}$ \cite{Datta2009,Neuling2012}) and do not directly elicit spikes, they can modulate spiking activity and lead to changes in cognitive processing, offering a range of possible clinical interventions \cite{Marshall2006,Berenyi2012,Herrmann2013}.  
How external fields lead to changes of the membrane voltage in single cells has been studied in detail \cite{Bikson2004,Radman2009,Deans2007}. However, their effects on population spike rate and the underlying mechanisms are largely unexplored.

Computational models of neurons exposed to electric fields offer a useful tool to gain a better understanding of these mechanisms. 
Multi-compartment models of neurons are well suited for corresponding investigations at the level of single cells and small circuits \cite{Tiganj2014} but are too complex for a purposeful application in large populations.
Single-compartment (point) neuron models of the integrate-and-fire (IF) type are well applicable to study the dynamics of large neuronal populations, due to their computational efficiency and analytical tractability \cite{Brunel2000}. However, typical IF model neurons lack the dendritic morphology required for a biophysical description of electric field effects.
Furthermore, even in the absence of an extracellular field, the dendritic morphology strongly shapes neuronal response properties \cite{Ostojic2015}.

In this contribution, we extend the popular class of IF point neuron models to quantitatively account for morphology dependent modulations of neural activity due to: (i) dendritic influences on the integration of synaptic inputs and (ii) the effects of extracellular electric fields.
Furthermore, we describe how oscillatory electric fields affect neuronal subthreshold and spiking activity and identify field-induced spike rate resonance. 
Specifically, we considered a canonical spatial pyramidal neuron model which consists of a somatic compartment and one (apical main) passive dendritic cable, and which is exposed to in-vivo like fluctuating synaptic input and an electric field. Based on that model we analytically derived an extension for the classical leaky and the refined exponential, \cite{Fourcaud2003}, IF point neuron models in order to exactly reproduce the subthreshold dynamics of the spatial model for arbitrary parametrizations. We then evaluated the extended IF models by quantitatively comparing their spiking activity with the spiking activity of the corresponding spatial model. Finally, we used these models to study the effects of an oscillating electric field (due to the presence of the dendritic cable) on the spike rate dynamics.
%

%
\section*{Results}

\begin{figure}[h!]
	\begin{center}
	\includegraphics[width=0.7\textwidth]{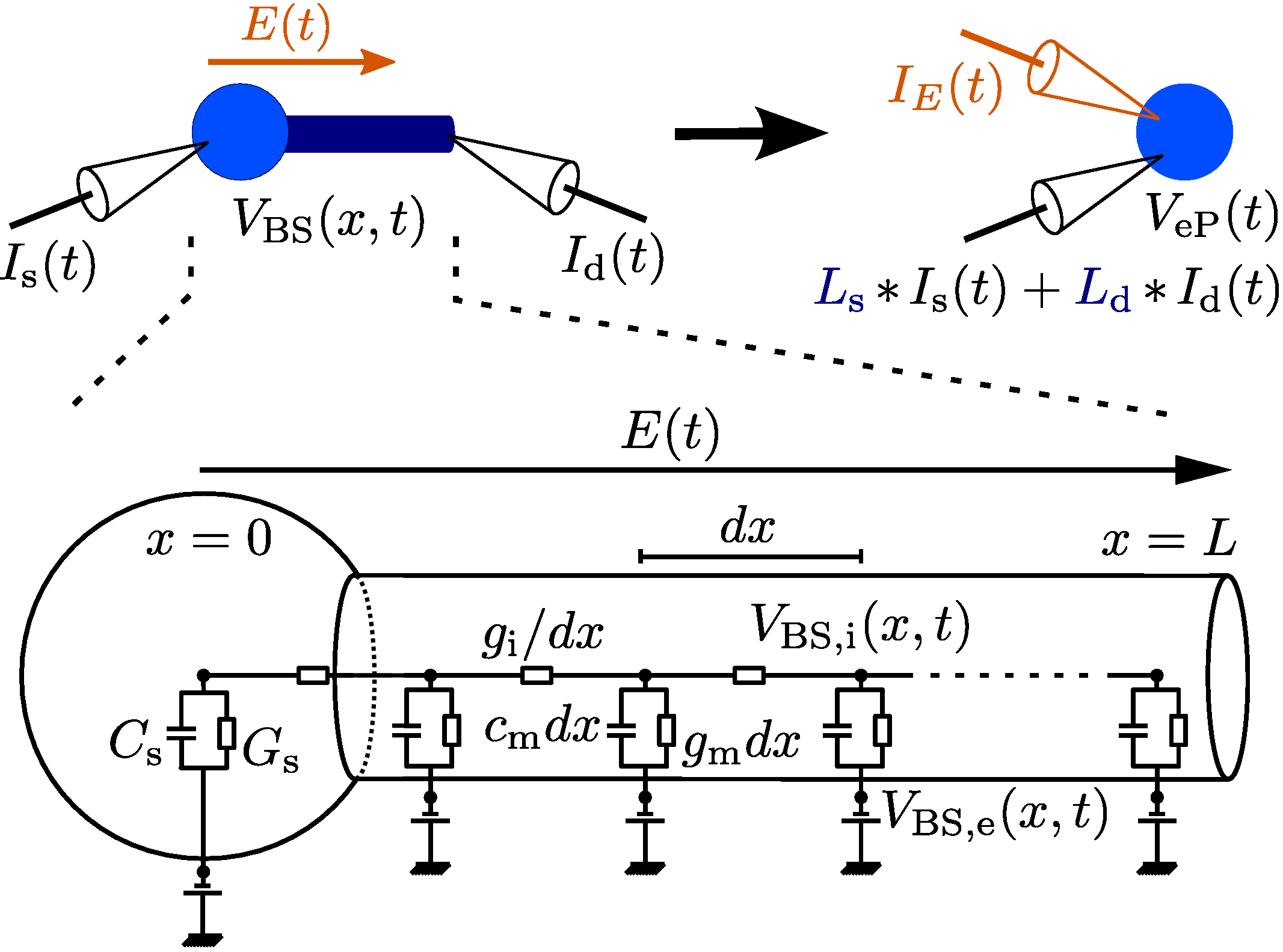}
	\end{center}
	\caption{{\bf Diagram of the extended point neuron model}
		\textit{Top:} Visualization of the ball-and-stick, BS, (\textit{left}) and the extended point, eP, (\textit{right}) neuron models. Both models receive synaptic input currents at the soma and the distal dendrite, $I_{\mathrm{s}}(t)$ and  $I_{\mathrm{d}}(t)$, and are exposed to an external electric field $ E(t) $.
		$L_\mathrm{s}(t)$ and $L_\mathrm{d}(t)$ denote the additional input filters describing the dendritic effects.
		$I_E(t)$ denotes the additional input current describing the field effect. 
		 $V_{\mathrm{eP}}(t)$ and $V_{\mathrm{BS}}(0,t)$ denote the corresponding membrane voltages (at the soma).
		\textit{Bottom:} Electrical circuit diagram for the subthreshold dynamics of the BS model.
		For a description of the parameters and their values see Table~\ref{table:BS_parameters}. 
	}
	\label{fig1}
\end{figure}

Our derivation of the extended point neuron model consists of two steps. 
We first calculate the somatic membrane voltage of a ball-and-stick (BS) model in response to subthreshold synaptic inputs at the soma and the distal dendrite
and to a time-varying, spatially homogeneous, extracellular electric field. This involves solving a generalized cable equation \cite{Rattay1986}.
Second, we seek to exactly reproduce this voltage response in the point neuron model by deriving additional model components (see Fig.~\ref{fig1}): two linear temporal filters, one for each input location, %
to be applied to the ``raw'' synaptic input and one additional input current to describe the field effect.
The model components are given in analytical form and depend on the parameters of the BS model and the electric field.
We refer to the model equipped with the new components as the extended point (eP) neuron model. 
We first derive this extension for the well-known leaky IF (LIF) neuron model, and present the extension adapted for the exponential IF (EIF) neuron model in a separate section.

\subsection*{Models}\label{sec:results_methods}
The BS neuron model consists of a lumped soma attached to a passive dendritic cable of length $ L $. The dynamics of its membrane voltage, when receiving synaptic inputs at the soma, $ I_{\mathrm{s}}(t) $, and the distal dendrite, $ I_{\mathrm{d}}(t) $, and when exposed to a spatially homogeneous external electric field, $ E(t) $, are governed by the cable equation:
\begin{equation}\label{eq:BSmodel}
c_{\mathrm{m}} \frac{\partial V_{\mathrm{BS}}}{\partial t} - g_{\mathrm{i}} \frac{\partial ^2 V_{\mathrm{BS}}}{\partial x ^ 2} + g_{\mathrm{m}} V_{\mathrm{BS}} = 0, \qquad 0<x<L, %
\end{equation}
subject to the boundary conditions:
\begin{align}
C_{\mathrm{s}} \frac{\partial V_{\mathrm{BS}}}{\partial t} 
- g_{\mathrm{i}}\frac{\partial V_{\mathrm{BS}}}{\partial x} + 
G_{\mathrm{s}} V_{\mathrm{BS}} - G_{\mathrm{s}} \Delta_{\mathrm{T}} 
e^{\tfrac{V_{\mathrm{BS}}-V_{\mathrm{T}}}{\Delta_{\mathrm{T}}}} &= 
I_{\mathrm{s}}(t) - g_{\mathrm{i}} E(t), \qquad x=0, \label{eq:BSmodel_BC_soma} \\ %
\frac{\partial V_{\mathrm{BS}}}{\partial x} &= \frac{ I_\mathrm{d}(t) }{ g_\mathrm{i} } +  E(t), \qquad x=L, \label{eq:BSmodel_BC_end} %
\end{align}
at the soma (x=0) and the end of the dendrite (x=L). $ V_{\mathrm{BS}} $ denotes the deviation of the membrane voltage from rest, $V_{\mathrm{rest}}$, 
$ V_{\mathrm{BS}}(x,t) := V_{\mathrm{BS,i}}(x,t)-V_{\mathrm{BS,e}}(x,t)-V_{\mathrm{rest}} $, where $ V_{\mathrm{BS,i}} $ and $ V_{\mathrm{BS,e}} $ are the intra- and extracellular potentials.
The effects of a spike are described by the IF-type reset condition for the soma:
\begin{equation}\label{eq:BSmodel_RC} 
\textrm{if }\, V_{\mathrm{BS}}(0,t) \ge V_{\mathrm{s}} \,\textrm{ then }\,
V_{\mathrm{BS}}(0,t) := V_{\mathrm{r}}
\end{equation}
and by a short refractory period of length $ T_{\mathrm{ref}} $ during which $ V_{\mathrm{BS}}(0,t) $ is clamped at the reset value $ V_{\mathrm{r}} $.
Spike times are defined by the times at which the somatic membrane voltage $ V_{\mathrm{BS}}(0,t) $ crosses the spike voltage value $ V_{\mathrm{s}} $ from below.
\textit{c}\textsubscript{m} denotes the membrane capacitance, \textit{g}\textsubscript{m} the membrane conductance, and \textit{g}\textsubscript{i} the internal (axial) conductance of a dendritic cable segment of unit length.
\textit{C}\textsubscript{s} and \textit{G}\textsubscript{s} are the somatic membrane capacitance and leak conductance.
The exponential term with threshold slope factor $\Delta$\textsubscript{T} and effective threshold voltage \textit{V}\textsubscript{T} approximates the somatic sodium current at spike initiation \cite{Fourcaud2003}.
For details see \nameref{sec:methods}.
 
In the proposed IF point neuron extension, that is, the eP model, the deviation of the membrane voltage, $ V_{\mathrm{eP}} $, from rest is governed by 
\begin{equation}\label{eq:Pmodel}
C_{\mathrm{eP}} \frac{dV_{\mathrm{eP}}}{dt} + G_{\mathrm{eP}} V_{\mathrm{eP}} 
- \alpha G_{\mathrm{eP}} \Delta_{\mathrm{T}} 
e^{\tfrac{V_{\mathrm{eP}}-V_{\mathrm{T}}}{\Delta_{\mathrm{T}}}} = 
[L_{\mathrm{s}} \ast I_{\mathrm{s}}](t)  + [L_{\mathrm{d}} \ast I_{\mathrm{d}}](t) + I_E(t),
\end{equation}
and by the reset condition: 
\begin{equation}\label{eq:Pmodel_RC} 
\textrm{if }\, V_{\mathrm{eP}} \ge V_{\mathrm{s}} \,\textrm{ then }\,
V_{\mathrm{eP}} := V_{\mathrm{r}}^\prime ,
\end{equation}
where $ V_{\mathrm{eP}} $ is clamped to $ V_{\mathrm{r}}^\prime $ for the duration of the refractory period $ T_{\mathrm{ref}} $ after every spike.
$ C_{\mathrm{eP}} $ and $ G_{\mathrm{eP}} $ are the membrane capacitance and leak conductance. 
The scaling factor $ \alpha $ ensures an equal membrane voltage response to the depolarizing current described by the exponential terms in both models (BS and eP).
We consider two versions of these models separately. First, we treat the LIF versions in detail, where we omit the exponential terms in Eqs.~{\ref{eq:BSmodel_BC_soma}} and {\ref{eq:Pmodel}}; specifically, by taking the limit $ \Delta_{\mathrm{T}} \to 0 $ (and setting $ V_{\mathrm{s}}=V_{\mathrm{T}} $). In the subsequent part we then consider the (full) EIF versions of the BS and eP models.
Below we explain in detail how the components of the point model extension are derived: the linear input filters $ L_{\mathrm{s}}(t) $, $ L_{\mathrm{d}}(t) $, the additional input current equivalent to the field effect,
$ I_E(t) $, and, in case of the (full) EIF type models, the scaling factor $ \alpha $. 
The analytical expressions of these model components are given in Eqs.~{\ref{eq:Ls}},~{\ref{eq:Ld}} and~{\ref{eq:IE_LIF}},~{\ref{eq:IE_LIF2}} (for the LIF case), and in Eqs.~{\ref{eq:Ld_EIF}}--{\ref{eq:B_EIF}} (for the EIF case).
To mimic the remaining depolarization along the dendritic cable after each spike, we choose an elevated reset voltage for all point neuron models: $ V_{\mathrm{r}}^\prime = (V_{\mathrm{r}} + V_{\mathrm{T}})/2 $.

For comparison we also use a point neuron model (of LIF and EIF type, respectively) without the extension, that is, $L_{\mathrm{s}}(t) =L_{\mathrm{d}}(t) = \delta (t)$ and $\alpha = 1$, and we fit the parameters of that model to best reproduce the activity of the BS model for equal synaptic inputs (details see below). We refer to this model as the P model. 
 
\subsection*{The somatic input filter for the LIF model}
We first consider the BS and eP model neurons of the LIF type (i.e, $ \Delta_{\mathrm{T}} \to 0 $, $ V_{\mathrm{s}}=V_{\mathrm{T}} $) receiving subthreshold synaptic input at the soma in the absence of an electric field ($ E(t) = 0 $, $ I_E(t) = 0 $, $ I_\mathrm{d}(t)=0 $). To avoid ambiguity we use the superscript $ I_\mathrm{s} $ for the membrane voltage variables in this case. %
The somatic membrane voltage response of the BS model (Eqs.~\ref{eq:BSmodel}--\ref{eq:BSmodel_BC_end}) %
can be calculated as (see \nameref{sec:methods})
\begin{equation}
\hat{V}_{\mathrm{BS}}^{ I_\mathrm{s} }(0,\omega) = \frac{\hat{I}_{\mathrm{s}}(\omega)}{C_{\mathrm{s}} \mathrm{i} \omega + G_{\mathrm{s}} + z(\omega) \,g_{\mathrm{i}} \, \mathrm{tanh}(z(\omega) L)}, \label{eq:BS_somatic_V_resp}
\end{equation}
\begin{equation}\label{eq:z}
z(\omega) = \sqrt{\frac{g_{\mathrm{m}} + \sqrt{g_{\mathrm{m}}^2 + \omega^2 c_{\mathrm{m}}^2}}{2g_{\mathrm{i}}}} +  \mathrm{sgn}(\omega) \mathrm{i} \sqrt{\frac{-g_{\mathrm{m}} + \sqrt{g_{\mathrm{m}}^2 + \omega^2 c_{\mathrm{m}}^2}}{2g_{\mathrm{i}}}},
\end{equation}
where $ \hat{.} $ indicates the temporal Fourier transform and $ \omega = 2\pi f $ denotes angular frequency. The somatic membrane voltage response of the eP model (Eq.~\ref{eq:Pmodel}) is given by
\begin{equation}
\hat{V}_{\mathrm{eP}}^{ I_\mathrm{s} }(\omega) = \frac{\hat{L}_{\mathrm{s}}(\omega) \hat{I}_{\mathrm{s}}(\omega)}{C_{\mathrm{eP}} \mathrm{i} \omega + G_{\mathrm{eP}}}. \label{eq:eP_somatic_V_resp}
\end{equation}
The dendritic filter $ L_{\mathrm{s}} $ required to exactly reproduce the somatic membrane voltage response of the BS model, i.e., $\hat{V}_{\mathrm{eP}}^{ I_\mathrm{s} }(\omega) = \hat{V}_{\mathrm{BS}}^{ I_\mathrm{s} }(0,\omega)$, must then be equal to ratio of the impedances of both models:
\begin{equation}\label{eq:Ls}
\hat{L}_{\mathrm{s}}(\omega) = \frac{C_{\mathrm{eP}} \mathrm{i} \omega + G_{\mathrm{eP}}}{C_{\mathrm{s}} \mathrm{i} \omega + G_{\mathrm{s}} + z(\omega) \,g_{\mathrm{i}} \, \mathrm{tanh}(z(\omega) L)},
\end{equation}
where $ z(\omega) $ is given by Eq.~\ref{eq:z}. In the following, we choose the membrane capacitance and conductance of the eP model to be equal to the corresponding somatic quantities of the BS model, $ C_{\mathrm{eP}}=C_{\mathrm{s}} $, $ G_{\mathrm{eP}} = G_{\mathrm{s}} $.
To see the necessity of the filter, let us consider the P model (no dendritic filter, $\hat{L}_{\mathrm{s}}(\omega)=1$) whose subthreshold response is given by
\begin{equation} 
\hat{V}_{\mathrm{P}}^{I_\mathrm{s}}(\omega) = \frac{\hat{I}_{\mathrm{s}}(\omega)}{C_{\mathrm{P}} \mathrm{i} \omega + G_{\mathrm{P}}}. \label{eq:P_somatic_V_resp}
\end{equation}
Because of the additional frequency-dependent term in the denominator of Eq.~\ref{eq:BS_somatic_V_resp} compared to Eq.~\ref{eq:P_somatic_V_resp}, it is not possible to adjust the parameters $ C_{\mathrm{P}}$ and $ G_{\mathrm{P}}$ of the P model such that $\hat{V}_{\mathrm{P}}^{ I_\mathrm{s} }(\omega) = \hat{V}_{\mathrm{BS}}^{ I_\mathrm{s} }(0,\omega)$ for all frequencies $ \omega $.
The somatic response of the BS model can only be approximated in this case.

\begin{figure}[!ht]
	\begin{center}
	\includegraphics[width=0.73\textwidth]{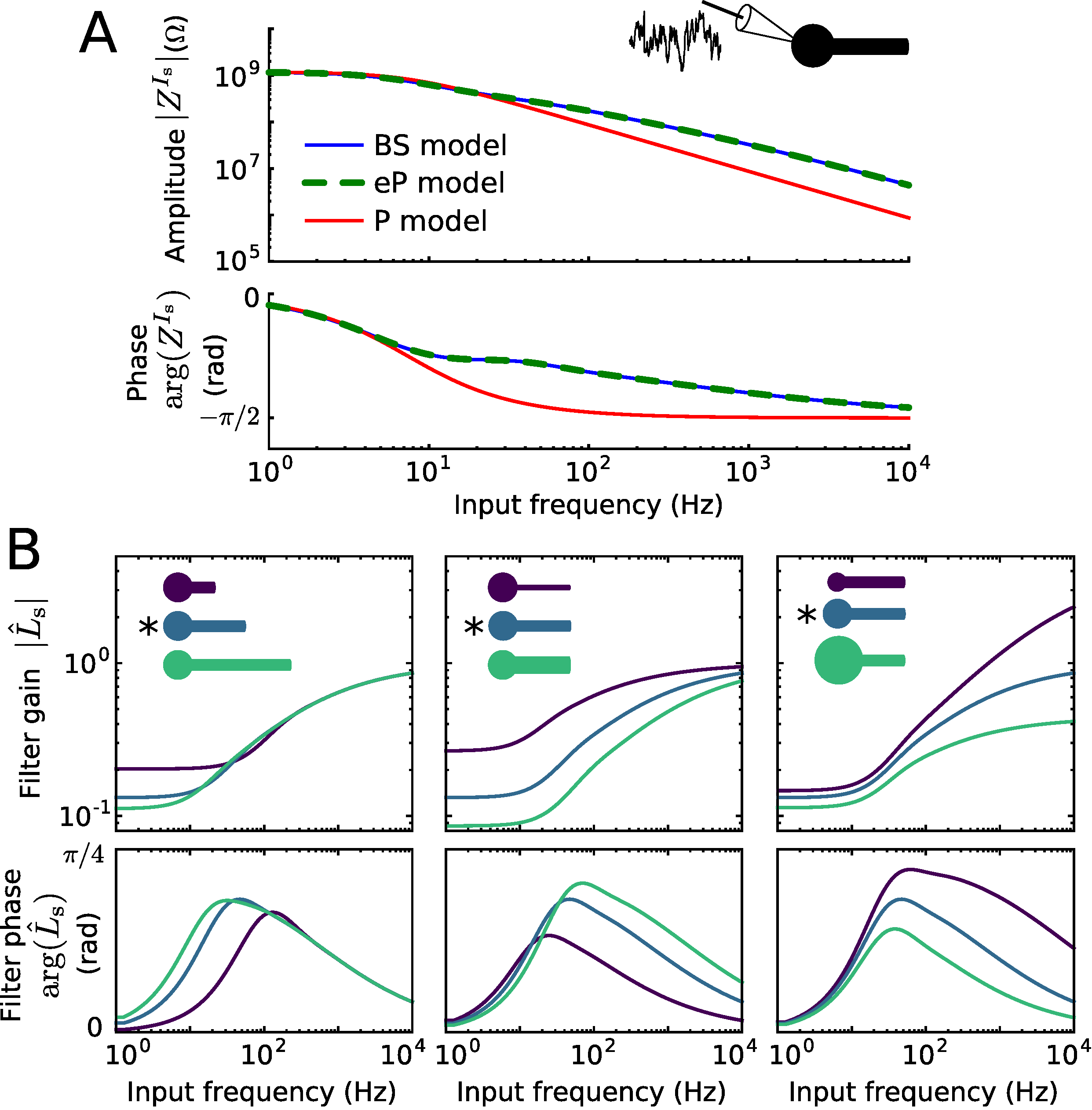}
	\end{center}
	\caption{{\bf Impedance and filters for somatic inputs} %
		A: Impedances $ Z_{\mathrm{BS}}^{ I_\mathrm{s} } $, $ Z_{\mathrm{eP}}^{ I_\mathrm{s} } $, and $ Z_{\mathrm{P}}^{ I_\mathrm{s} } $ of the three neuron models as a function of input frequency. 
		B: Gain and phase of the input filter  $\hat{L}_\mathrm{s}$ as a function of input frequency. 
		The neuronal morphology varied as indicated, in terms of dendritic cable length (350~$ \mu \mathrm{m} $, 700~$ \mu \mathrm{m} $, 1050~$ \mu \mathrm{m} $), cable diameter (0.6~$ \mu \mathrm{m} $, 1.2~$ \mu \mathrm{m} $, 1.8~$ \mu \mathrm{m} $) and soma diameter (5~$ \mu \mathrm{m} $, 10~$ \mu \mathrm{m} $, 15~$ \mu \mathrm{m} $). 
		$\ast$ indicates the default parameter values.
		For all other parameter values used see Table~\ref{table:BS_parameters}.
	}
	\label{fig3}
\end{figure}

Fig.~\ref{fig3}A shows the impedances, 
 $ Z_{\mathrm{m}}^{ I_\mathrm{s} }(\omega) := \hat{V}_{\mathrm{m}}^{ I_\mathrm{s} }(\omega)/\hat{I}_{\mathrm{s}}(\omega)$, $\mathrm{m} \in \{ \mathrm{BS}|_{x=0},\mathrm{eP},\mathrm{P} \}$,
of the three neuron models for an example set of parameter values for the BS model.
The two parameters of the P model ($ C_{\mathrm{P}}$ and $G_{\mathrm{P}} $) were determined by matching the steady-state somatic voltage, $ Z_{\mathrm{P}}^{ I_\mathrm{s} }(0) = Z_{\mathrm{BS}}^{ I_\mathrm{s} }(0) $, and minimizing the mean squared distance between $ Z_{\mathrm{P}}^{ I_\mathrm{s} } $ and $ Z_{\mathrm{BS}}^{ I_\mathrm{s} } $ over the visualized range of input frequencies. %
The impedance of the eP model matches the impedance of the BS model exactly while the impedance of the P model deviates substantially, in particular for larger frequencies. 

Fig.~\ref{fig3}B-D show the amplitudes and phases of the input filter $\hat{L}_\mathrm{s}(\omega)$ for various sets of parameters for the BS morphology.
$\hat{L}_\mathrm{s}(\omega)$ is always a high-pass filter, which attenuates the somatic inputs at lower and amplifies them at higher frequencies.
This effect is more pronounced for a larger dendritic and a smaller somatic compartment. It becomes stronger with increasing ratio of dendritic over somatic size. Nevertheless, the filter does not differ qualitatively for changes in neuron morphology.

\begin{figure}[!ht]
	\begin{center}
	\includegraphics[width=0.8\textwidth]{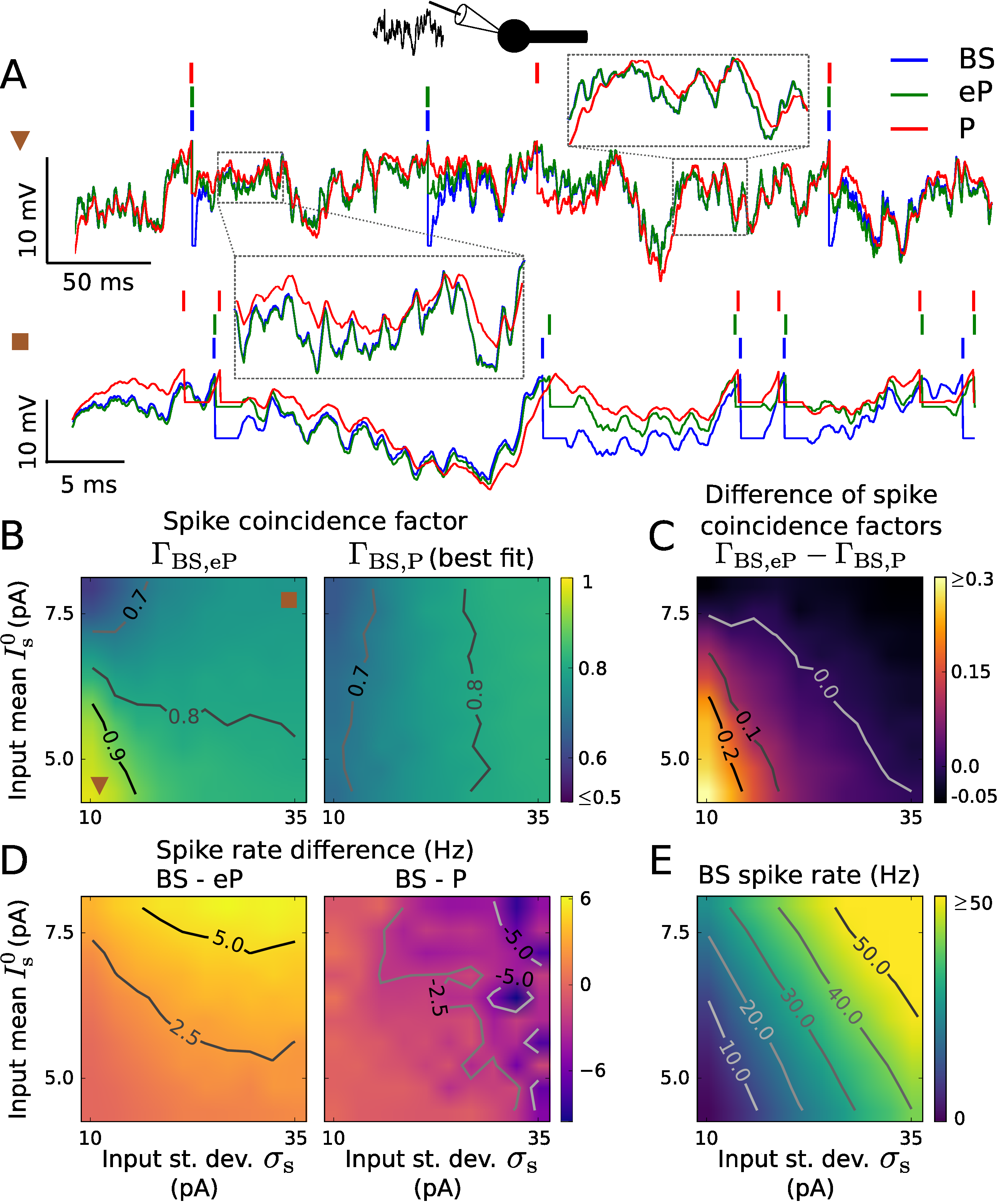}
	\end{center}
	\caption{{\bf Reproduction of spiking activity for somatic inputs using LIF type models}
		A: Membrane voltage traces of the BS (blue), eP (green) and P (red) neuron models in response to a weak ($I_\mathrm{s}^0=4.68$~pA, $\sigma_\mathrm{s}=11.94 $~pA, top) and a strong input current ($I_\mathrm{s}^0=7.69$~pA, $\sigma_\mathrm{s}=33.34$~pA, bottom). The parameter values of the P model were tuned independently to maximize the coincidence factor $\Gamma_{\mathrm{BS, P}}$ for each set of input parameters. %
		B: Coincidence factor for the BS and eP model spike trains, $\Gamma_{\mathrm{BS, eP}}$ (left), and for the BS and P model spike trains, $\Gamma_{\mathrm{BS, P}}$ (right) as a function of input mean $ I_\mathrm{s}^0 $ and standard deviation $ \sigma_\mathrm{s} $. C: Difference $\Gamma_{\mathrm{BS, eP}} - \Gamma_{\mathrm{BS, P}}$ between the coincidence factors shown in B.
		D: Spike rate difference of the BS and eP models (left) and of the BS and P models (right) as a function of $ I_\mathrm{s}^0 $ and $ \sigma_\mathrm{s} $. 
		E: Spike rate of the BS neuron model.
		The input parameters used in A are indicated in B. Results presented in B-E are averages over 6 noise realizations. The parameter values of the BS model are listed in Table~\ref{table:BS_parameters}.
	}
	\label{fig4}
\end{figure}

We next compare how well the point neuron models eP and P reproduce the spiking activity of the BS model neuron. 
For this purpose we consider an in vivo-like fluctuating synaptic input current $ I_\mathrm{s}(t) $ described by an Ornstein-Uhlenbeck process (see \nameref{sec:methods}). The model outputs are compared over a range of values for the input mean $I_\mathrm{s}^0$ and standard deviation $\sigma_\mathrm{s}$. 
The parameter values of the P model were adjusted to best reproduce the spike train of the BS model (see \nameref{sec:methods} for details). 
Fig.~\ref{fig4}A displays the time series of the somatic membrane voltage of the three models in response to the same input currents -- a weak (small $I_\mathrm{s}^0$, $\sigma_\mathrm{s}$) and a strong current (large $I_\mathrm{s}^0$, $\sigma_\mathrm{s}$).
For both input currents, the eP model well reproduces the somatic voltage dynamics of the BS model.  Consequently, the spike times are also well reproduced.
There is, however, a mismatch between the voltage traces during short periods (of less than approximately 10~ms duration) after spikes have occurred. This discrepancy is a result of the remaining dendritic depolarization after a spike has occurred in the BS model, which is only approximated by the elevated reset voltage $ V_{\mathrm{r}}^\prime $ (see section~\nameref{sec:results_methods} above) in the point neuron models.  
In comparison, the P model performs worse in reproducing the BS membrane voltage dynamics, particularly the fast fluctuations are poorly recovered. This is expected from the mismatch in the impedance for high frequencies (cf. Fig.~\ref{fig3}A). 

In Fig.~\ref{fig4}B-E we compare spiking activity in terms of spike coincidences and spike rates for a wide range of input parameters. 
We used the spike coincidence measure $\Gamma$ which quantifies the similarity between two spike trains for a given precision of 3 ms (see~\nameref{sec:methods}).
The maximum value of 1 indicates an optimal match, i.e., spike times always coincide, a value of 0 corresponds to pure chance, 
i.e., the degree of coincidences for two Poisson spike trains with equal rates.
The P model was fitted to the BS model for each input (in terms of $I_\mathrm{s}^0$, $\sigma_\mathrm{s}$) separately. %
The parameters of the eP model, on the other hand, are constant and do not depend on the input at all.
The eP model very accurately reproduces the BS spike times for small spike rates ($\Gamma \geq 0.9$ for small $I_\mathrm{s}^0$ and $\sigma_\mathrm{s}$), see Fig.~\ref{fig4}B,E. This performance decreases only slightly for increasing $\sigma_\mathrm{s}$ (noise dominated input) and somewhat stronger for increasing $I_\mathrm{s}^0$ (mean dominated input). 
Generally, $ \Gamma $ decreases with increasing spike rates. This can be attributed to the transient periods after spikes during which the dendritic cable is still loaded and the membrane voltages of both neuron models deviate. Those periods do not depend on the spike rate and therefore have a stronger deteriorating effect when the interspike intervals are smaller. 
In addition, when $\sigma_\mathrm{s}$ is small the model neurons spike repetitively in a rather clock-like manner, with comparable rate but most likely out of phase due to mismatches caused by the membrane voltage resets. %
This helps understand the rather low values of $ \Gamma $ for mean dominated inputs.
The spike rate of the BS model is also reproduced quite well by the eP model, which underestimates it only slightly (Fig.~\ref{fig4}D).
Spike coincidence and spiking rate reproduction of the eP model can be improved even further by additionally tuning the reset voltage $ V_{\mathrm{r}}^\prime $ using $ \Gamma $ or the spike rate distance as a cost function. %
The P model, in comparison, is substantially worse in reproducing the spike times at small spike rates and only slightly better than the eP model for large spike rates (Fig.~\ref{fig4}B,C).
The spike rate of the BS model is slightly overestimated by the P model (Fig.~\ref{fig4}D).
Even though the parameters of the P model were optimized in an input-dependent manner the eP model leads to an improved reproduction of the BS spiking activity overall.

In summary, the dendritic cable implements a high pass filter for inputs at the soma. Due to the derived filter for somatic inputs, the eP model -- without having fitted any of its parameters -- well reproduces the BS model dynamics for subthreshold and suprathreshold inputs. 
Notably, the computation time required for the
BS model was at least 25 times that of the eP model, using measurements on a single
core of a desktop computer.

\subsection*{The distal input filter for the LIF model} %
\begin{figure}[!ht]
	\begin{center}
	\includegraphics[width=0.8\textwidth]{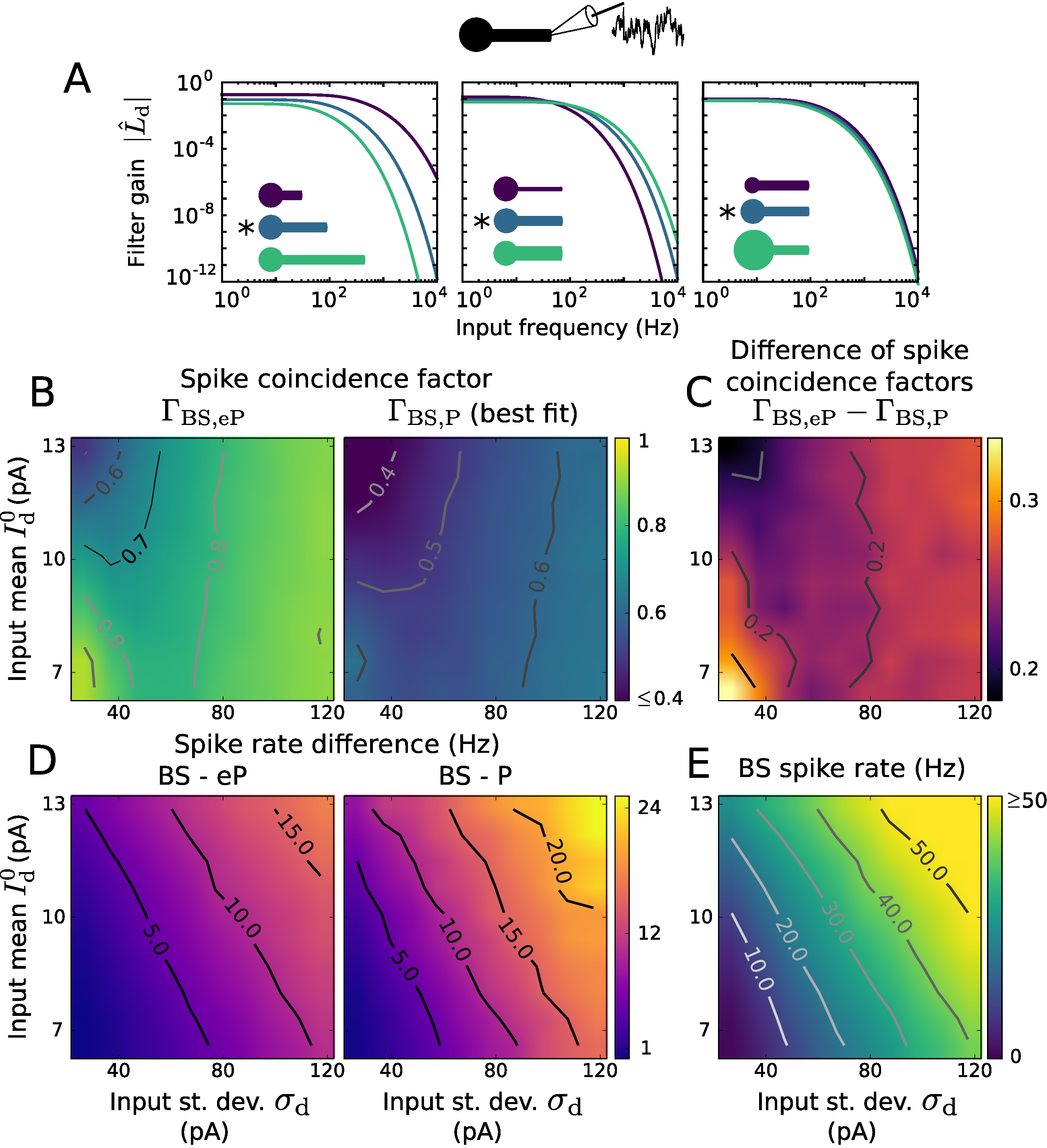}
	\end{center}
	\caption{{\bf Distal input filter and reproduction of spiking activity using LIF type models}
				A: Gain and phase of the input filter  $\hat{L}_\mathrm{d}$ as a function of frequency.
				The neuronal morphology varied as indicated, in terms of dendritic cable length (350~$ \mu \mathrm{m} $, 700~$ \mu \mathrm{m} $, 1050~$ \mu \mathrm{m} $), cable diameter (0.6~$ \mu \mathrm{m} $, 1.2~$ \mu \mathrm{m} $, 1.8~$ \mu \mathrm{m} $) and soma diameter (5~$ \mu \mathrm{m} $, 10~$ \mu \mathrm{m} $, 15~$ \mu \mathrm{m} $). 
				$\ast$ indicates the default parameter values.
				B: Coincidence factor for the BS and eP model spike trains, $\Gamma_{\mathrm{BS, eP}}$ (left), and for the BS and P model spike trains, $\Gamma_{\mathrm{BS, P}}$ (right) as a function of input mean $ I_\mathrm{d}^0 $ and standard deviation $\sigma_\mathrm{d} $.
				C: Difference  $\Gamma_{\mathrm{BS, eP}} - \Gamma_{\mathrm{BS, P}}$ between the coincidence factors shown in B.
				D: Spike rate difference of the BS and eP models (left) and of the BS and P models (right) as a function of $ I_\mathrm{d}^0 $ and $ \sigma_\mathrm{d} $. 
				E: Spike rate of the BS neuron model. Results presented in B-E are averages over 6 noise realizations. The default parameters values of the BS model are listed in Table~\ref{table:BS_parameters}.
	}
	\label{fig_Id1}
\end{figure}

We next consider subthreshold synaptic input at the distal dendrite instead of somatic input, but otherwise the same setup as in the previous section. Here we use superscipt $ I_\mathrm{d} $ for the membrane voltage variables to better distinguish from the previous scenario.
The somatic membrane voltage response of the BS model can be expressed as (see \nameref{sec:methods})
\begin{equation}
\hat{V}_{\mathrm{BS}}^{ I_\mathrm{d} }(0,\omega) = \frac{\hat{I}_{\mathrm{d}}(\omega) \, \mathrm{sech}(z(\omega) L)}{C_{\mathrm{s}} \mathrm{i} \omega + G_{\mathrm{s}} + z(\omega) \, g_{\mathrm{i}} \, \mathrm{tanh}(z(\omega) L)}, \label{eq:BS_somatic_V_resp_Id}
\end{equation}
where $ z(\omega) $ is given by Eq.~\ref{eq:z}. In order to reproduce that voltage response using the eP model, for which $ \hat{V}_{\mathrm{eP}}^{ I_\mathrm{d} }(\omega) = \hat{L}_{\mathrm{d}}(\omega) \hat{I}_{\mathrm{d}}(\omega) / (C_{\mathrm{eP}} \mathrm{i} \omega + G_{\mathrm{eP}}) $ (cf. Eq.~\ref{eq:eP_somatic_V_resp}), we obtain
\begin{equation}\label{eq:Ld}
\hat{L}_{\mathrm{d}}(\omega) = \frac{(C_{\mathrm{eP}} \mathrm{i} \omega + G_{\mathrm{eP}}) \, \mathrm{sech}(z(\omega) L)}{C_{\mathrm{s}} \mathrm{i} \omega + G_{\mathrm{s}} + z(\omega) \, g_{\mathrm{i}} \, \mathrm{tanh}(z(\omega) L)}.
\end{equation}
As in the previous section, we choose $ C_{\mathrm{eP}}=C_{\mathrm{s}} $, $ G_{\mathrm{eP}} = G_{\mathrm{s}} $.
In contrast to the somatic input filter $L_\mathrm{s}$ the filter $L_\mathrm{d}$ for distal inputs exhibits low pass properties for various BS morphologies, see Fig.~\ref{fig_Id1}A. The shape of this filter is largely independent of the soma size.
Compared to the attenuation of low frequency in case of somatic input, the filter gain for high frequency dendritic input is much lower. This results in a stronger filtering effect for dendritic inputs than for somatic inputs.

An evaluation of the distal input filter in terms of reproduction of BS spiking activity ($ \Gamma $ and rates) is shown in Fig.~\ref{fig_Id1}B-E for a range of input mean $I_\mathrm{d}^0$ and standard deviation $\sigma_\mathrm{d}$ values. For comparison we used the P model (without filter) whose parameters were tuned to best reproduce the spike train of the BS model for each input (i.e., ($I_\mathrm{d}^0$, $\sigma_\mathrm{d}$)-pair) separately.
The eP model very accurately reproduces the BS spike times for small spike rates ($\Gamma \geq 0.9$ for small $I_\mathrm{d}^0$ and $\sigma_\mathrm{d}$). The accuracy drops somewhat as $I_\mathrm{d}^0$ increases, which can be explained as in the previous section. Interestingly, the performance does not deteriorate with increasing spike rate in general; it remains high if the noise intensity $\sigma_\mathrm{d}$ is sufficiently strong ($\Gamma \geq 0.8$ for $\sigma_\mathrm{d} \geq 80$ pA, independent of $I_\mathrm{d}^0$ in the considered range).
The spike rate of the BS model is somewhat underestimated by the eP model (Fig.~\ref{fig_Id1}D). It should be noted that the spike rate reproduction could be substantially improved by an increased reset voltage value $ V_{\mathrm{r}}^\prime $, as the remaining dendritic depolarization after spikes is more pronounced in case of distal input compared to somatic input.
The computational speed-up of the eP model here is the same as in the previous section. 
The P model, in comparison, is less accurate across all inputs (Fig.~\ref{fig_Id1}B-D), even though its parameters depend on the input. 

In summary, the dendritic cable implements a low pass filter for inputs at the distal dendrite, and due to the corresponding derived filter the eP model reproduces the BS model dynamics for subthreshold and suprathreshold inputs much better than the P model.

\subsection*{Effect of an extracellular electric field on the neuronal dynamics}
We now consider an extracellular electric field -- in addition to the synaptic inputs -- to which the neuron is exposed to. We characterize the effects of that field on the subthreshold somatic membrane voltage and spiking dynamics of the BS neuron and we determine an explicit expression for the additional input current of the extended point neuron model to reproduce these effects.
The electric fields we are interested in are oscillatory, spatially uniform on the neuronal scale and weak such as induced by transcranial brain stimulation \cite{Bikson2012a}. In the following, we consider a field with amplitude $ E_1 $ and angular frequency $ \varphi $,
\begin{equation} \label{eq:electric_field}
E(t) = -\frac{\partial V_\mathrm{BS,e}}{\partial x}(t) = E_1 \sin (\varphi t).
\end{equation}
Recall that $V_\mathrm{BS,e}(x,t)$ is the extracellular potential. The BS subthreshold somatic membrane voltage response to this field, $V_{\mathrm{BS}}^E (0,t)$, is determined by Eqs.~\ref{eq:BSmodel}--\ref{eq:BSmodel_BC_end}.
Using the temporal Fourier transform the solution can be expressed analytically as 
\begin{equation}  \label{eq:BS_somatic_V_resp_E}
\hat{V}_{\mathrm{BS}}^E (0,\omega) = \frac{\hat{E}(\omega) g_{\mathrm{i}} \,[\mathrm{sech}(z(\omega) L) -1]}{ C_{\mathrm{s}} \mathrm{i} \omega + G_{\mathrm{s}} + z(\omega) \, g_{\mathrm{i}} \, \mathrm{tanh}(z(\omega) L)},
\end{equation}
where $ z(\omega) $ is given by Eq.~\ref{eq:z} (see \nameref{sec:methods}). Note, that we again neglect the exponential current in this section (LIF case, %
but see next section for the EIF case). 
In the time domain this yields
\begin{align}
V_{\mathrm{BS}}^E (0,t) &= | A(\varphi) | \, \sin \! \big(\varphi t + \arg(A(\varphi))\big), \\ 
A(\varphi) &= \frac{ E_1 g_{\mathrm{i}} \,[\mathrm{sech}(z(\varphi) L) -1]}{ C_{\mathrm{s}} \mathrm{i} \varphi + G_{\mathrm{s}} + z(\varphi) \, g_{\mathrm{i}} \, \mathrm{tanh}(z(\varphi) L)}, \label{eq:A}
\end{align}
where $\arg(x)$ denotes the argument of the complex number $x$. 
The overall subthreshold response in presence of the electric field and synaptic input can be decomposed as %
\begin{equation}
\hat{V}_{\mathrm{BS}}(0,\omega) = \hat{V}_{\mathrm{BS}}^{I_{\mathrm{s}}}(0,\omega) + \hat{V}_{\mathrm{BS}}^{I_{\mathrm{d}}}(0,\omega) + \hat{V}_{\mathrm{BS}}^E(0,\omega),
\end{equation} 
with $ \hat{V}_{\mathrm{BS}}^{I_{\mathrm{s}}}(0,\omega) $, $ \hat{V}_{\mathrm{BS}}^{I_{\mathrm{d}}}(0,\omega) $ and $ \hat{V}_{\mathrm{BS}}^E(0,\omega) $ given by Eqs.~\ref{eq:BS_somatic_V_resp}, \ref{eq:BS_somatic_V_resp_Id} and \ref{eq:BS_somatic_V_resp_E}. For the eP model, on the other hand, we have
\begin{equation}
\hat{V}_{\mathrm{eP}}(\omega) = \frac{\hat{L}_{\mathrm{s}}(\omega)\hat{I}_{\mathrm{s}}(\omega) + \hat{L}_{\mathrm{d}}(\omega)\hat{I}_{\mathrm{d}}(\omega) +  \hat{I}_E(\omega)}{C_{\mathrm{eP}} \mathrm{i} \omega + G_{\mathrm{eP}}}. \label{eq:P_somatic_V_resp_E}
\end{equation} 
To guarantee an equal subthreshold response in both models, i.e., $ \hat{V}_{\mathrm{eP}}(\omega) = \hat{V}_{\mathrm{BS}}(0,\omega) $, we obtain the following expression for the additional input current,
\begin{align}\label{eq:IE_LIF}
I_E(t) &= | B(\varphi) | \, \sin \! \big(\varphi t + \arg(B(\varphi))\big), \\
B(\varphi) &= \frac{E_1 g_{\mathrm{i}} (C_{\mathrm{eP}} \mathrm{i} \varphi + G_{\mathrm{eP}}) [\mathrm{sech}(z(\varphi) L) -1]}{C_{\mathrm{s}} \mathrm{i} \varphi + G_{\mathrm{s}} + z(\varphi) \, g_{\mathrm{i}} \, \mathrm{tanh}(z(\varphi) L)}, \label{eq:IE_LIF2}
\end{align}
where we set $ C_{\mathrm{eP}}=C_{\mathrm{s}} $ and $ G_{\mathrm{eP}} = G_{\mathrm{s}} $ (as in the previous sections). 
It should be noted that these results are not restricted to sinusoidal field variations, as considered here, and can be easily adjusted for any time-varying or constant description of the electric field using its Fourier transform.

\begin{figure}[!ht]
	\begin{center}
	\includegraphics[width=0.73\textwidth]{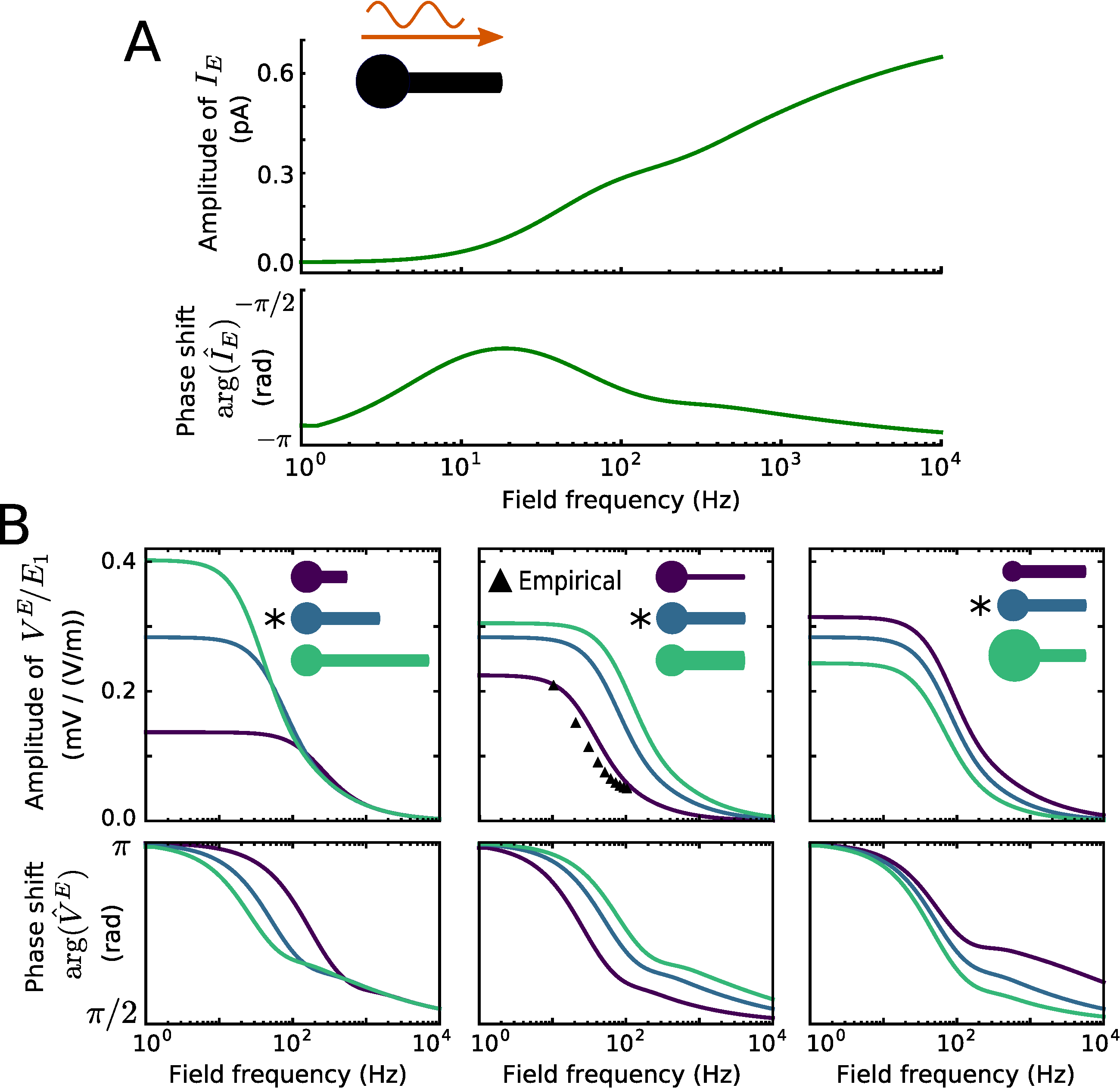}
	\end{center}
	\caption{{\bf Input current equivalent to the field effect and somatic sensitivity} %
		A: Input current $I_E$ to reproduce the effect of a 1~V/m field in the eP model. Its amplitude and phase shift relative to the field as a function of field frequency.
		B: Neuron sensitivity to the field, i.e., the ratio between its somatic membrane voltage amplitude and the field amplitude, and phase shift between the oscillatory membrane voltage and the field.
		The neuronal morphology varied as indicated, in terms of dendritic cable length (350~$ \mu \mathrm{m} $, 700~$ \mu \mathrm{m} $, 1050~$ \mu \mathrm{m} $), diameter (0.6~$ \mu \mathrm{m} $, 1.2~$ \mu \mathrm{m} $, 1.8~$ \mu \mathrm{m} $) and soma diameter (5~$ \mu \mathrm{m} $, 10~$ \mu \mathrm{m} $, 15~$ \mu \mathrm{m} $). 
		$\blacktriangle$ indicate values obtained from electrophysiological recordings of rat hippocampal pyramidal cells \cite{Deans2007}.
				$\ast$ indicates the default parameter values.
				For all other parameter values used see Table~\ref{table:BS_parameters}.	}
	\label{fig6}
\end{figure}

The equivalent input current $ I_E(t) $ as well as the somatic subthreshold sensitivity to the field, $ |A(\varphi)|/E_1$ and the phase shift between oscillating membrane voltage and field, $ \arg(A(\varphi)) $, with $ A(\varphi) $ from Eq.~\ref{eq:A}, are shown in Fig.~\ref{fig6}.
Interestingly, the amplitude of $ I_E(t) $ increases with increasing field frequency (Fig.~\ref{fig6}A), while the sensitivity decreases (Fig.~\ref{fig6}B).
The sensitivity curve changes quantitatively, but not qualitatively, with varying neuronal morphology (Fig.~\ref{fig6}B). Specifically, its dependence on the field frequency becomes more pronounced with increasing ratio of dendritic size over somatic one. The cable length has the strongest impact in this respect. 
Notably, the morphology parameters can be adjusted such that the sensitivity curve well matches with empirical results obtained from rat hippocampal pyramidal cells in vitro.
The phase shift between the somatic membrane voltage and field oscillations also depends on the field frequency. It exhibits an anti-phase relation for slow oscillations, and decreases with increasing frequency (Fig.~\ref{fig6}B).

We next assess how the electric field affects spiking activity for a range of field frequencies using the BS and eP models.
For that purpose, we simulated both model neurons subject to the field and noisy synaptic input at the soma or at the distal dendrite. The synaptic drive alone is strong enough to cause stochastic spiking with rate $ r_0 $.  
The oscillatory field leads to an oscillatory spike rate modulation quantified as $ r_1(\varphi) \sin(\varphi t + \psi(\varphi)) $ around the constant baseline spike rate $ r_0 $ (see \nameref{sec:methods} for details). 
Note that this spike rate modulation measure is related to the frequently used spike field coherence measure.

\begin{figure}[!ht]
	\begin{center}
	\includegraphics[width=0.9\textwidth]{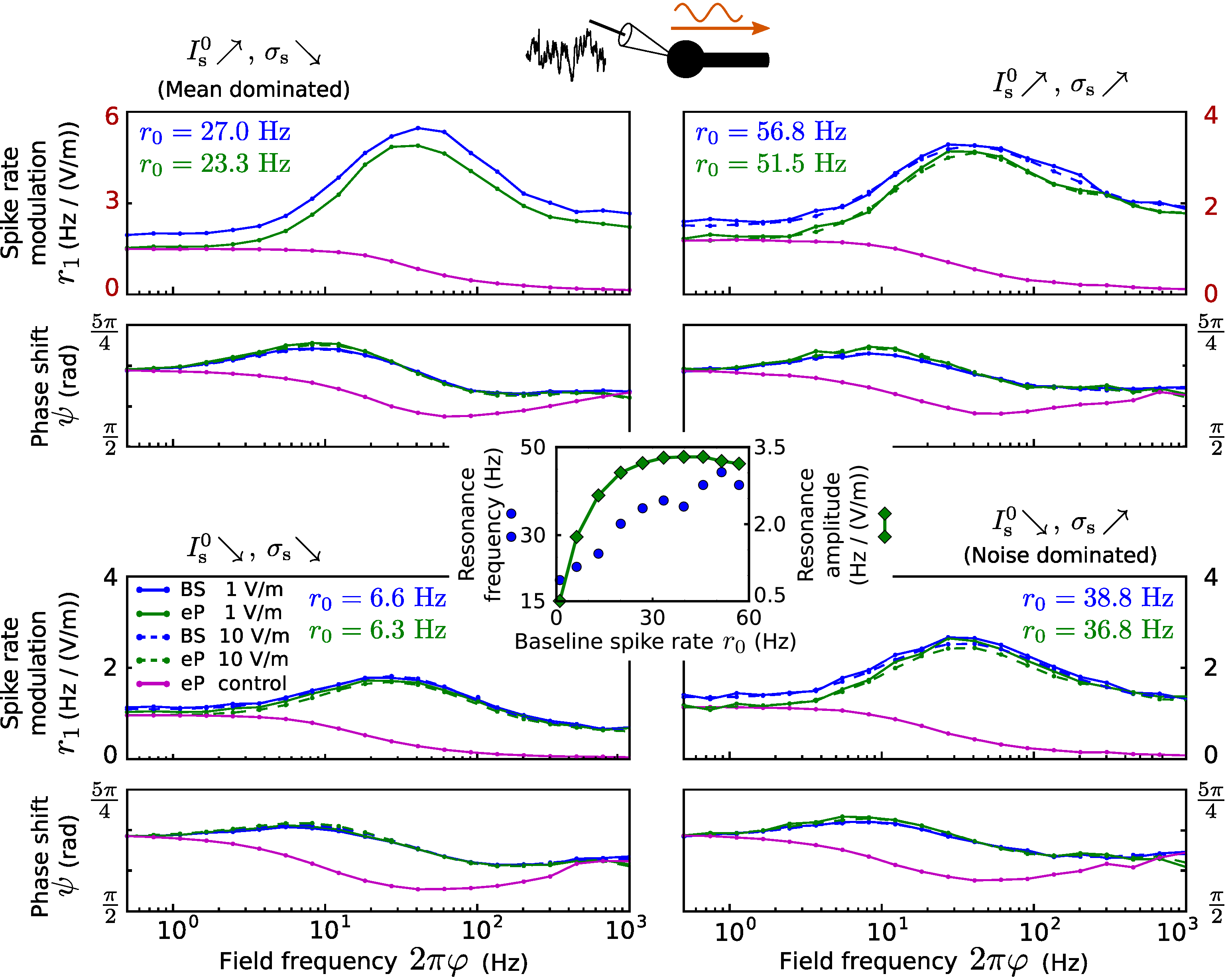}
	\end{center}
	\caption{{\bf Spike rate modulation due to an electric field for somatic inputs}
		Top/bottom, left/right: Spike rate modulation of the BS (blue) and the eP (green) models due to an oscillating electric field as a function of its frequency, for different field amplitudes ($ E_1=1$~V/m, solid lines;  $ E_1=10$~V/m, dashed lines) and somatic inputs: $ I_\mathrm{s}^0 =7.69 $~pA,  $\sigma_\mathrm{s} =11.94 $~pA (top left), $ I_\mathrm{s}^0 =7.69 $~pA, $\sigma_\mathrm{s} =33.34 $~pA (top right), $ I_\mathrm{s}^0 =4.68 $~pA, $\sigma_\mathrm{s} =11.94 $~pA (bottom left), and $ I_\mathrm{s}^0 =4.68 $~pA, $\sigma_\mathrm{s} =33.34 $~pA (bottom right).
		Magenta lines show the spike rate modulation of the eP model for which $ I_E $ was given by $ I_E(t) = I_1 \sin(\varphi t + \phi) $ with constant amplitude $ I_1 = |B(0.5/(2\pi))|$, phase shift $\phi = \arg\left(B(0.5/(2\pi))\right)$, $B$ from Eq.~\ref{eq:IE_LIF2} and $ E_1=10 $~V/m.
		Note the different amplitude scales in the top panel.
		Results for larger field amplitude ($ E_1=10$~V/m) are not displayed for the mean driven regime (top right), because spike rate modulation amplitudes exceeded the baseline rate in that case, which impedes the modulation quantification procedure (see \nameref{sec:methods}).
		Center: Resonance frequency $\mathrm{argmax}(r_1) $ and amplitude $ \mathrm{max}(r_1) $ of the spike rate modulation of the eP model as a function of baseline spike rate $ r_0 $, which was changed by simultaneously increasing ($I_\mathrm{s}^0$, $ \sigma_\mathrm{s} $) from (4.25 pA, 8.89 pA) to (8.12 pA, 36.40 pA).
	}	
	\label{fig7}
\end{figure}

The amplitude $ r_1 $ and phase shift $ \psi $ of the spike rate modulation for various somatic inputs (in terms of $ I_\mathrm{s}^0 $ and $ \sigma_\mathrm{s} $), a range of field oscillation frequencies and two field strengths are shown in Fig.~\ref{fig7}.
The eP model well reproduces the spike rate dynamics of the BS model exposed to the field for all considered field and input parameter values. The amplitude $ r_1 $ increases linearly with increasing field magnitude $ E_1 $.
In contrast to the subthreshold sensitivity to the field (cf. Fig.~{\ref{fig6}}B), the spike rate modulation exhibits a clear resonance in the beta and gamma frequency bands across the different inputs. In other words, the spike rate oscillations are strongest for field oscillations of that frequency range. The amplitude peak is more pronounced for stronger inputs and most prominent when the input is dominated by its mean (large $I_\mathrm{s}^0$, small $\sigma_\mathrm{s}$). 
This resonance amplitude rapidly increases with increasing baseline spike rate -- by increasing both, mean and standard deviation of the background input from small values -- and saturates at about $ r_0 = 30 $~Hz (Fig.~{\ref{fig7}}, center). The resonance frequency shifts rather gradually from the beta to the gamma range as the baseline spike rate increases from a few spikes per second to about 60~Hz.
The phase shift $\psi$ varies around $\pi$, depending on the input and field frequency. Note that $\psi = \pi$ implies that the probability of spiking is largest at the trough of the field oscillation. This results from the orientation of the field, which, in case of $ E(t) = E_0 > 0 $, induces a (hyper-)polarized somatic membrane voltage.

To examine the importance of the specific shape of $ I_E(t) $, we also considered an alternative sinusoidal input current $ I_E(t) = I_1 \sin(\varphi t + \phi) $ for the eP model. Note that the amplitude and phase shift of that current are constant across different field frequencies.
Using that current, the typical resonance of the spike rate modulation due to the field cannot even roughly be reproduced (Fig.~\ref{fig7}).

\begin{figure}[!ht]
	\begin{center}
		\includegraphics[width=0.9\textwidth]{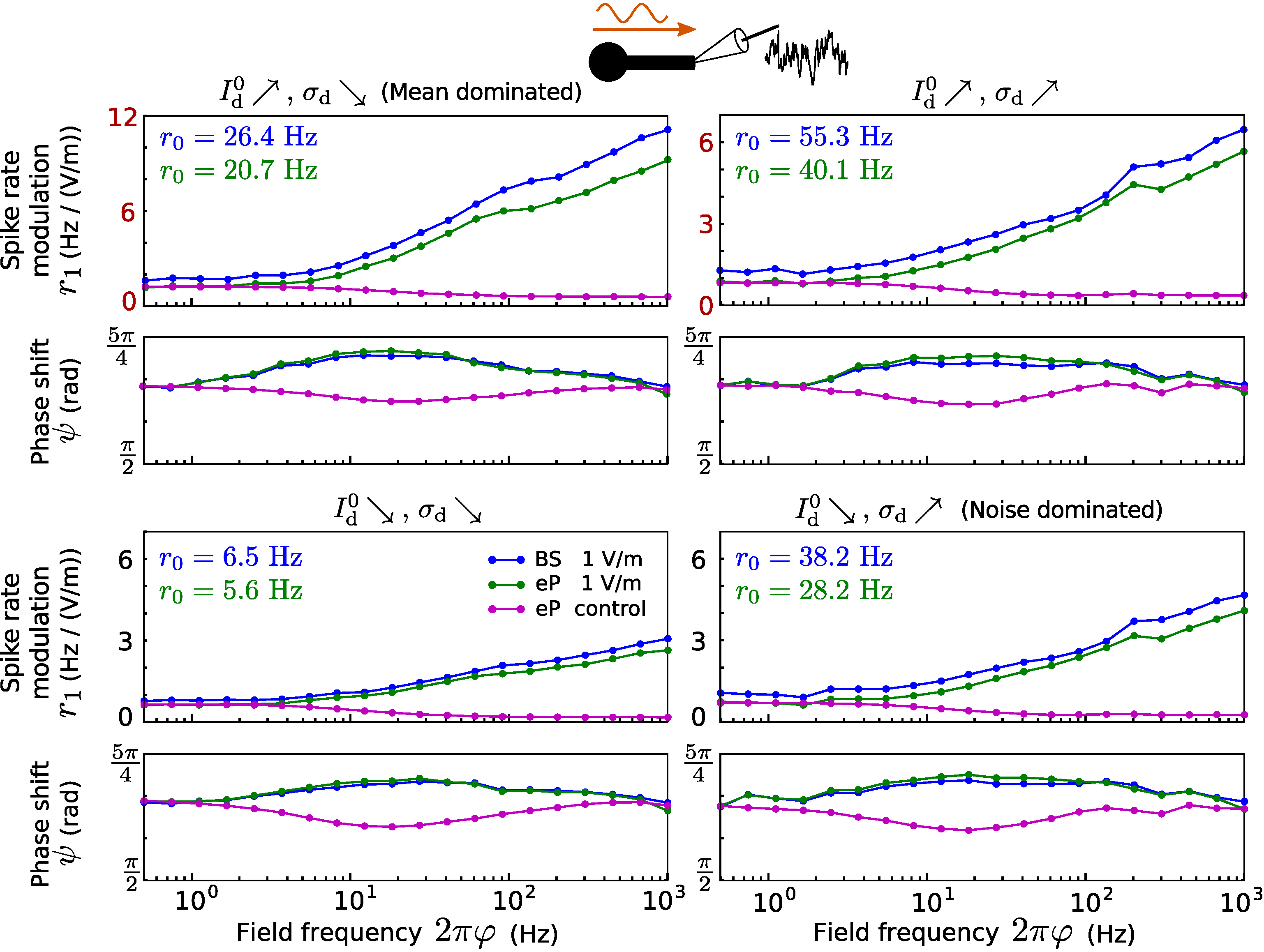}
	\end{center}
	\caption{{\bf Spike rate modulation due to an electric field for distal dendritic inputs}
		Spike rate modulation of the BS (blue) and the eP (green) models due to an oscillating electric field as a function of its frequency, for different distal dendritic inputs: $ I_\mathrm{d}^0 =12.44 $~pA,  $\sigma_\mathrm{d} = 33.04 $~pA (top left), $ I_\mathrm{d}^0 =12.44 $~pA, $\sigma_\mathrm{d} =111.2 $~pA (top right), $ I_\mathrm{d}^0 =7.03 $~pA, $\sigma_\mathrm{d} =33.04 $~pA (bottom left), and $ I_\mathrm{d}^0 =7.03 $~pA, $\sigma_\mathrm{d} =111.2 $~pA (bottom right).
		Magenta lines show the spike rate modulation of the eP model for which $ I_E $ was given by $ I_E(t) = I_1 \sin(\varphi t + \phi) $ with constant amplitude $ I_1 = |B(0.5/(2\pi))|$ and phase shift $\phi = \arg\left(B(0.5/(2\pi))\right)$ with $B$ from Eq.~\ref{eq:IE_LIF2} and $ E_1=10 $~V/m.
		Note the different amplitude scales in the upper panel.
	}	
	\label{fig_fRates_dend}
\end{figure}

Let us now inspect spike rate modulation due to the field in presence of distal dendritic inputs instead of somatic ones. In Fig.~\ref{fig_fRates_dend} the results are shown for various distal inputs (in terms of $ I_\mathrm{d}^0 $ and $ \sigma_\mathrm{d} $). Interestingly, for all considered distal dendritic inputs, spike rate modulation amplitudes increase monotonically with the field frequency over the whole considered range (up to 1~kHz, see Discussion for an explanation). %
 Similarly as for somatic inputs, modulation is strongest for mean dominated (large $I_\mathrm{d}^0$, small $\sigma_\mathrm{d}$) distal inputs, and the phase shift $\psi$ varies around $\pi$. Overall, the eP model well reproduces the modulation observed in the BS model.

\subsection*{Extension for EIF model neurons}\label{subsec:EIF}
In the previous sections, we considered only capacitive and leak currents through the neuronal membrane; the model extension presented there applies to the LIF type model neurons.
Here, we consider the BS and eP models described by Eqs.~\ref{eq:BSmodel}--\ref{eq:BSmodel_BC_end},~\ref{eq:Pmodel} without neglecting the exponential term, that approximates the voltage dependent sodium current at spike initiation. That is, we derive and evaluate the model extension for model neurons of the EIF type.  

To derive the required model components $ L_\mathrm{s}(t)$, $ L_\mathrm{d}(t)$, $ \alpha $ and $ I_E(t) $ we linearize the exponential terms in Eqs.~\ref{eq:BSmodel_BC_soma} and \ref{eq:Pmodel} around a baseline voltage value $ V_0 $ and then proceed similarly as above. Specifically, we calculate the subthreshold somatic membrane voltage response of the BS model, using the (temporal) Fourier transform, and obtain four response components: 
$ \hat{V}_{\mathrm{BS}}(0,\omega) = \hat{V}_{\mathrm{BS}}^{I_{\mathrm{s}}}(0,\omega) + \hat{V}_{\mathrm{BS}}^{I_{\mathrm{d}}}(0,\omega) + \hat{V}_{\mathrm{BS}}^{\Delta_{\mathrm{T}}}(0,\omega) + \hat{V}_{\mathrm{BS}}^E(0,\omega) $, 
where $ V_{\mathrm{BS}}^{I_{\mathrm{s}}} $, $ V_{\mathrm{BS}}^{I_{\mathrm{d}}} $ and $ V_{\mathrm{BS}}^E $ denote the voltage response components to $ I_{\mathrm{s}} $, $ I_{\mathrm{d}} $ and $ E $, respectively, and the additional term $ V_{\mathrm{BS}}^{\Delta_{\mathrm{T}}} $ is due to the (linearized) exponential term.
These four voltage response components are given by the explicit expressions Eqs.~\ref{eq:V_BS_Is}--\ref{eq:V_BS_X} in the \nameref{sec:methods} section.
For the eP model, on the other hand, we can also calculate the subthreshold membrane voltage response in the Fourier domain, $ \hat{V}_{\mathrm{eP}}(\omega) $, given by Eq.~\ref{eq:V_eP}.
By requiring equal subthreshold responses, 
$\hat{V}_{\mathrm{eP}}(\omega) = \hat{V}_{\mathrm{BS}}(0,\omega) $, we obtain the following explicit expressions for the components $ L_\mathrm{s}$, $ L_\mathrm{d}$, $ \alpha $ and $ I_E $, considering the electric field defined in Eq.~{\ref{eq:electric_field}}:
\begin{figure}[h!]
	\begin{center}
		\includegraphics[width=0.8\textwidth]{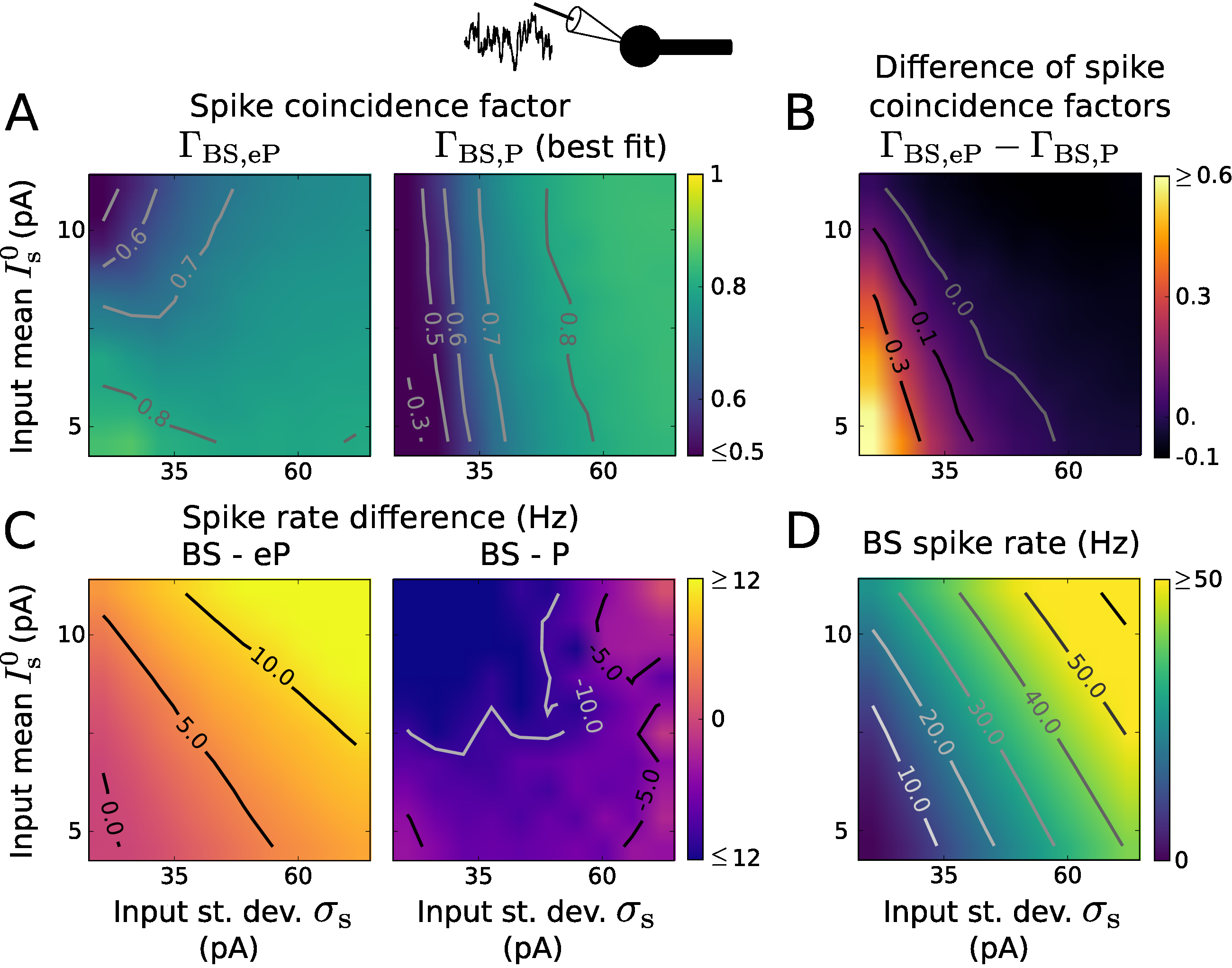}
	\end{center}
	\caption{{\bf Reproduction of spiking activity for somatic inputs using EIF type models}
		A: Coincidence factor for the BS and eP model spike trains, $\Gamma_{\mathrm{BS, eP}}$ (left), and for the BS and P model spike trains, $\Gamma_{\mathrm{BS, P}}$ (right) as a function of input mean $ I_\mathrm{s}^0 $ and standard deviation $ \sigma_\mathrm{s} $. The parameter values of the P model were optimized to maximize $\Gamma_{\mathrm{BS, P}}$ for each input (i.e., ($ I_\mathrm{s}^0 $, $ \sigma_\mathrm{s} $)-pair) independently. 
		B: Difference $\Gamma_{\mathrm{BS, eP}} - \Gamma_{\mathrm{BS, P}}$ between the coincidence factors shown in B.
		C: Spike rate difference of the BS and eP models (left) and of the BS and P models (right) as a function of $ I_\mathrm{s}^0 $ and $ \sigma_\mathrm{s} $. 
		D: Spike rate of the BS neuron model.
		Results presented in A-D are averages over 6 noise realizations. The parameter values of the BS model are listed in Table~\ref{table:BS_parameters}.
	}
	\label{fig8}
\end{figure}
\begin{align}\label{eq:Ld_EIF}
\hat{L}_{\mathrm{s}}(\omega) &= \frac{C_{\mathrm{eP}} \mathrm{i} \omega + G_{\mathrm{eP}} \Big(1- \alpha e^{\tfrac{V_0-V_{\mathrm{T}}}{\Delta_{\mathrm{T}}}} \Big)}{C_{\mathrm{s}} \mathrm{i} \omega + G_{\mathrm{s}}\Big(1-e^{\tfrac{V_0-V_{\mathrm{T}}}{\Delta_{\mathrm{T}}}} \Big) + z(\omega) \, g_{\mathrm{i}} \, \mathrm{tanh}(z(\omega) L)}, \\
\hat{L}_{\mathrm{d}}(\omega) &= \frac{\Big[ C_{\mathrm{eP}} \mathrm{i} \omega + G_{\mathrm{eP}} \Big(1- \alpha e^{\tfrac{V_0-V_{\mathrm{T}}}{\Delta_{\mathrm{T}}}} \Big)\Big] \,\mathrm{sech}(z(\omega) L)}{C_{\mathrm{s}} \mathrm{i} \omega + G_{\mathrm{s}}\Big(1-e^{\tfrac{V_0-V_{\mathrm{T}}}{\Delta_{\mathrm{T}}}} \Big) + z(\omega) \, g_{\mathrm{i}} \, \mathrm{tanh}(z(\omega) L)}, \label{eq:Ld2_EIF}\\
\alpha &= \frac{G_{\mathrm{s}}}{G_{\mathrm{s}} + \mathrm{tanh}(L/\lambda) \, g_{\mathrm{i}}/\lambda},
\end{align}
\begin{align}
I_E(t) &= | B(\varphi) | \, \sin \! \big(\varphi t + \arg(B(\varphi))\big), \\
\label{eq:B_EIF}
B(\varphi) &= \frac{E_1 g_{\mathrm{i}}  \Big[C_{\mathrm{eP}} \mathrm{i} \varphi + G_{\mathrm{eP}} \Big(1- \alpha e^{\tfrac{V_0-V_{\mathrm{T}}}{\Delta_{\mathrm{T}}}} \Big) \Big] [ \mathrm{sech}(z(\varphi) L) -1 ]}{C_{\mathrm{s}} \mathrm{i} \varphi + G_{\mathrm{s}}\Big(1-e^{\tfrac{V_0-V_{\mathrm{T}}}{\Delta_{\mathrm{T}}}} \Big) + z(\varphi) \, g_{\mathrm{i}} \, \mathrm{tanh}(z(\varphi) L)},
\end{align}
where $ z(\omega) $ is given by Eq.~\ref{eq:z}.
The scaling factor $\alpha$ guarantees that the voltage response component caused by the exponential term, $ V_{\mathrm{BS}}^{\Delta_{\mathrm{T}}} $, is reproduced. In other words, $\alpha$ ensures that the spike initiation current, described by the exponential term, leads to an equal steady state in both models.
Note that the two filters for EIF neurons and those for LIF neurons depend on input frequency in qualitatively the same way (by comparing Eqs.~{\ref{eq:Ld_EIF}} and {\ref{eq:Ld2_EIF}} with Eqs.~{\ref{eq:Ls}} and {\ref{eq:Ld}}). 

We assessed the reproduction of BS spiking activity by the extended EIF model for somatic inputs using the spike coincidence factor $ \Gamma $ and estimated spike rates (Fig.~\ref{fig8}).
Here again the parameter values of the P model were adjusted to maximize $\Gamma_{\mathrm{BS, P}}$ for each input separately. The range of input parameter values was chosen to obtain similar spike rates as in Fig.~\ref{fig4}. 
Despite the linearization in the derivation, the eP model achieves a correct reproduction of the BS spike trains ($\Gamma \geq 0.7$ for a wide range of input parameters). In particular, $\Gamma_{\mathrm{BS, eP}}$ is large for small spike rates (small $I_\mathrm{s}^0$ and $\sigma_\mathrm{s}$) and decreases for increasing $ I_\mathrm{s}^0 $ (towards mean dominated input), see Fig.~\ref{fig8}A,D. 
The eP model tends to underestimate the firing rate of the BS model (Fig.~\ref{fig8}C). This discrepancy in the rate could be reduced by optimizing the point model reset voltage, $ V_{\mathrm{r}}^\prime $, to better account for the remaining dendritic cable depolarization in the BS model.
Similarly, an improved performance of the eP model in terms of spike train reproduction could be achieved by tuning this reset voltage.
The P model, on the other hand, rather poorly reproduces the BS spiking dynamics for small input noise intensity ($\Gamma \leq 0.6$ for $\sigma_\mathrm{s} \leq 30$ pA, see Fig.~\ref{fig8}A).
Overall, also in presence of the exponential term the eP model clearly outperforms the simpler P model for small spike rates ($\Gamma_{\mathrm{BS,eP}} - \Gamma_{\mathrm{BS,P}} \geq 0.3$ for small $I_\mathrm{s}^0$ and $\sigma_\mathrm{s}$) and achieves similar performance for high spiking activity (Fig.~\ref{fig8}B).

\begin{figure}[ht!]
	\begin{center}
		\includegraphics[width=0.9\textwidth]{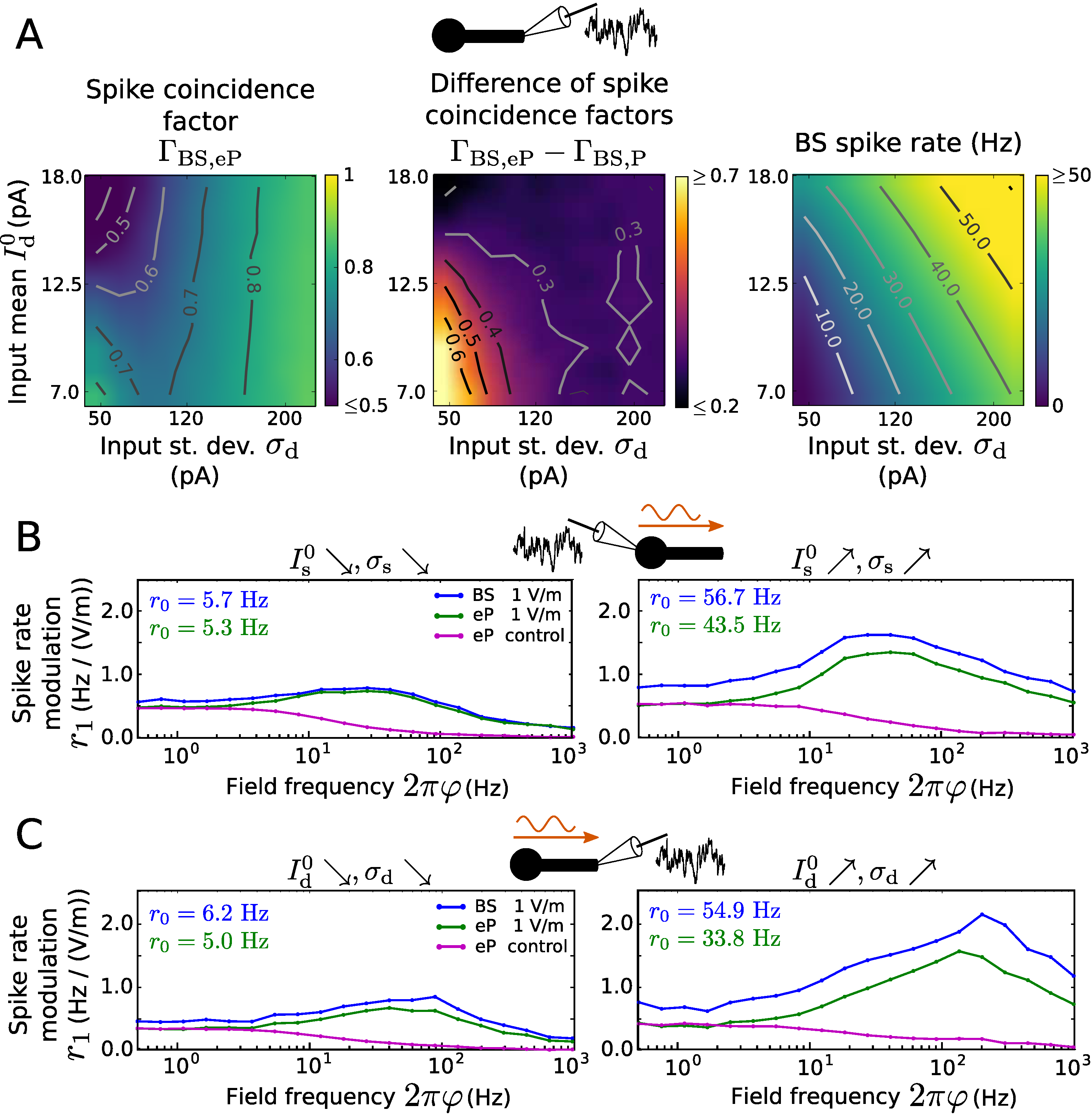}
	\end{center}
	\caption{{\bf Reproduction of spiking activity for dendritic inputs and spike rate modulation due to an electric field using EIF type models }
		A: Coincidence factor for the BS and eP model spike trains, $\Gamma_{\mathrm{BS, eP}}$ (left) as a function of input mean $ I_\mathrm{s}^0 $ and standard deviation $ \sigma_\mathrm{s} $.
		Difference $\Gamma_{\mathrm{BS, eP}} - \Gamma_{\mathrm{BS, P}}$ between the coincidence factors obtained with the eP and the P models (center).
		The parameter values of the P model were optimized to maximize $\Gamma_{\mathrm{BS, P}}$ for each input (i.e., ($ I_\mathrm{s}^0 $, $ \sigma_\mathrm{s} $)-pair) independently. 
		Spike rate of the BS neuron model (right).
		Results are averages over 6 noise realizations. The parameter values of the BS model are listed in Table~\ref{table:BS_parameters}.
		B:		Spike rate modulation of the BS (blue) and the eP (green) models due to an oscillating electric field as a function of its frequency, for different distal somatic inputs: $ I_\mathrm{s}^0 = 5.05$~pA,  $\sigma_\mathrm{s} = 24.08$~pA (left), $ I_\mathrm{s}^0 = 10.61$~pA, $\sigma_\mathrm{s} = 68.21$~pA (right).
		Magenta lines show the spike rate modulation of the eP model for which $ I_E $ was given by $ I_E(t) = I_1 \sin(\varphi t + \phi) $ with constant amplitude $ I_1 = |B(0.5/(2\pi))|$ and phase shift $\phi = \arg\left(B(0.5/(2\pi))\right)$ with $B$ from Eq.~\ref{eq:IE_LIF2} and $ E_1=10 $~V/m.
		C: Same as B for dendritic synaptic input instead of somatic one: $ I_\mathrm{d}^0 = 7.56$~pA,  $\sigma_\mathrm{d} = 57.73 $~pA (left), $ I_\mathrm{d}^0 = 16.73$~pA, $\sigma_\mathrm{d} = 203.41$~pA (right).
	}
	\label{fig9}
\end{figure}

The reproduction of spiking activity of the BS model was also assessed for distal dendritic inputs. The range of input parameters ($I_\mathrm{d}^0 $ and  $\sigma_\mathrm{d}$) was adjusted to obtain similar BS spike rates as for the LIF case.
The eP model performs well, in particular for small spike rates or sufficiently strong noise intensity; its performance decreases in the mean driven regime (Fig.~{\ref{fig9}}A). On the contrary the P model fails to reproduce the BS spiking activitiy (see Fig.~A in \nameref{sec:supps} for more details).

In summary, the somatic and distal dendritic input filters obtained for EIF neurons are qualitatively similar to the ones obtained for LIF neurons. The eP model, in contrast to the P model, well reproduces the BS model dynamics for subthreshold and suprathreshold inputs -- also for the EIF case.
 
Spike rate modulations due to an oscillatory electric field using EIF type model neurons for synaptic background input at the soma or distal dendrite are displayed in Fig.~{\ref{fig9}} (see also Fig.~B and Fig.~C in \nameref{sec:supps} for additional parameter values of the background input).
Similarly to the LIF case, spike rate modulation amplitudes do not decrease monotonically with the field frequency.
For somatic background input, we find spike rate resonance in the beta and gamma frequency range, similarly as shown by LIF type models. However, in case of distal dendritic input, EIF neurons exhibit resonance peaks in the high gamma frequency band, in contrast to LIF neurons, whose resonance frequency is substantially larger (see Discussion for an explanation). For both input locations the spike rate modulations shown by the BS model are well reproduced by the eP model and resonance amplitudes are stronger for large spike rates (i.e., large $I_\mathrm{s}^0 $, $\sigma_\mathrm{s}$ and large $I_\mathrm{d}^0 $, $\sigma_\mathrm{d}$, respectively).

%
%
%

%

\section*{Discussion}
In this contribution we presented an extension for IF point model neurons to accurately reflect the filtering of synaptic inputs caused by the presence of a dendrite and the effects of weak, oscillatory electric fields on neuronal activity. Based on a canonical BS neuron model,
we analytically derived additional components for LIF and EIF point neuron models to exactly reproduce the subthreshold voltage dynamics of the spatially extended BS neuron. 

These new components consist of (i) two linear filters applied to synaptic inputs depending on their location (soma or distal dendrite) and (ii) an additional input current quantifying the field effect on the membrane voltage.
The EIF point model requires an additional scaling parameter to accurately match the BS voltage dynamics.
Exhaustive evaluations for suprathreshold in-vivo like fluctuating inputs demonstrated that the BS spiking activity is well reproduced by the extended point neuron model in both cases (LIF and EIF).
Optimizing the parameters of the standard LIF and EIF models without the derived extension components, however, does not suffice to adequately reproduce the BS model dynamics.

Due to their computational efficiency the extensions of the point neuron models are well suited for application in large networks to investigate, for example, the effects of neuronal morphology and electrical fields on neuronal spiking activity at the population level.
Additionally, our methodological results serve as a building block to derive mean-field descriptions for the collective (spike rate) dynamics of large coupled populations \cite{Brunel2000,Richardson2009,Augustin2013}, \cite[chapter~4.2]{Ladenbauer2015}. An implementation of the presented models using Python (for the eP model) and NEURON (for the BS model) is freely available at \url{https://github.com/nigroup/IF_extension}.

Below, we summarize our results on the obtained input filters and the field effects on neuronal dynamics.  

\subsection*{Synaptic input filtering due to the dendrite}

We have demonstrated that synaptic input is integrated at the soma in distinct ways due to the presence of the dendrite, depending on the input site.
Distal dendritic input is low-pass filtered (cf. Fig.~\ref{fig_Id1}A), in accordance with previous results \cite{Koch2004}, whereas somatic input is high-pass filtered (cf. Fig.~\ref{fig4}B). %
The latter effect is consistent with recent measurements from Purkinje cells and with theoretical results \cite{Ostojic2015} which show a similar change in somatic impedance due to the presence of a dendritic tree (Fig.~4 in \cite{Ostojic2015}, in comparison with Fig.~\ref{fig3}A here).
Consequently, the presence of a dendrite can lead to an enhanced neuronal spiking response to high-frequency somatic inputs \cite{Ostojic2015}, which may be further amplified by the dendritic effect on the sharpness of spikes at the axon initial segment \cite{Eyal2014}. The derived IF model extension enables efficient analyses of the BS spike rate response to modulations of the input current -- which are, however, not within the scope of this paper.

There are two different strategies for taking into account complex neuron morphologies in models while keeping numerical simulation computationally efficient.
One option is to reduce the number of compartments while retaining important properties of the dendritic tree \cite{Pinsky1994}.
Alternatively, one can extend point neuron models with temporal kernels which are calibrated to reproduce the somatic membrane voltage response to synaptic inputs as observed in complex morphological cells \cite{Dayan2001,Jolivet2004}.
Our approach is of the latter type, with the advantage that the temporal kernels (filters) are  analytically derived from the underlying morphological BS model.

A similar extension for point model neurons to reproduce dendritic input integration of model cells with complex morphology has been recently proposed in \cite{Wybo2013}. Using the Green's function formalism a synapse model was developed, whose computational complexity practically allows for only a small number of synaptic input locations. Based on the BS model we were able to derive input filters for point model neurons using only the Fourier transform (without having to rely on the Green's function) and these filters are simple to implement. 

We have demonstrated that our extended model outperforms the simpler point neuron model in terms of spike train reproduction. Overall, it performs well for suprathreshold inputs, particularly in case of distal inputs and for somatic inputs that are not too strong. That performance could be further enhanced by optimizing the reset voltage to better reflect the remaining dendritic membrane depolarization in the BS model after each spike, as was mentioned previously.

In our study we have considered passive dendrites. Nonlinear (spike-generating) currents along the dendrite, which cause nonlinear synaptic input integration \cite{Migliore2002,Zhou2013,Zhang2013}, could be incorporated using our approach in a ``quasi-active'' framework \cite{Koch2004}. This would involve solving the cable equation with linearized nonlinear components, similarly as for the exponential terms used here (EIF case).

\subsection*{Effects of weak electric fields on neuronal activity}

We investigated in detail the effects of a spatially homogeneous, oscillating, weak electric field, as induced by transcranial electrical stimulation, on the activity of the BS neuron.
Such a one-dimensional spatial (cable plus soma) model provides a good approximation for neurons with elongated (apical) dendrites exposed to a uniform extracellular electrical field as long as the dendritic (apical main) cable is not substantially smaller than its electrotonic length~\cite[chapter~2.5]{Malik2011}.
Following the somatic doctrine \cite{Bikson2012a}, we focused on the effects of the field that are due to the polarization of the membrane voltage at the soma.
We analytically calculated the subthreshold voltage response, whose properties are in accordance with electrophysiological observations: the response magnitude scales linearly with the field amplitude \cite{Bikson2004}, as shown by the sensitivity in Fig.~\ref{fig6}. This sensitivity is of the same order of magnitude as that measured in pyramidal cells \cite{Deans2007}, i.e., around 0.30~mm for low frequency fields, and decreases with increasing field frequency in a morphology dependent manner \cite{Radman2009}.
For non-uniform electric fields, e.g., as generated by point source stimulation, however, the sensitivity can be roughly constant for frequencies up to at least 100~Hz \cite{Anastassiou2011}. Interestingly, such a behavior can also be observed for a uniform field in case of a rather short dendritic cable (cf. Fig.~{\ref{fig6}B}).

While polarization effects due to direct current fields have been extensively studied \cite{Cartee1992,Plonsey1988,Tranchina1986}, the effects of time-varying fields are less well understood. The response of the subthreshold membrane voltage to time-varying fields has been calculated in \cite{Monai2010} for a finite dendritic cable with leaky currents at one end, and in \cite{Anastassiou2010} for a spatially non-uniform field. 
Using a one-dimensional cable model \cite{Malik2011} showed that the electrotonic length is a key quantity that determines the neuronal subthreshold response to an electric field. Specifically, elongated neurons are less sensitive to high frequency fields than compact ones.
How the voltage response to an input current at a particular location along the cable depends on input frequency is largely determined by the membrane time constant. In case of an electrical field, however, which corresponds to symmetrical stimulation at both ends of the cable, the voltage response is also strongly influenced by currents flowing through the low-resistant intracellular medium. This results in an enhanced high frequency response to an extracellular field when compared to an input current~\cite[chapter~5]{Malik2011}.

Nevertheless, a somatic compartment was not considered in these studies. Using the BS model we have shown that the relative size of the soma compared to the dendritic cable substantially affects the neuronal sensitivity to the field.

Further, we found frequency-dependent spike rate modulation (and hence, spike field coherence) caused by the electric field. Unlike neuronal subthreshold sensitivity, spike rate modulation amplitude did not decrease with the field frequency and its precise relationship to field frequency depended on the synaptic input location. Spike rate modulation exhibited a clear resonance in the beta and gamma frequency bands in presence of only somatic inputs (cf. Figs.~\ref{fig7}, B in \nameref{sec:supps}), whereas for only distal dendritic inputs, spike rate modulation amplitudes are strongest at much larger frequencies (cf. Figs.~\ref{fig_fRates_dend}, C in \nameref{sec:supps}).
This can be linked to a theoretical result showing that the response of single-compartment model neurons to high frequency inputs is stronger for larger autocorrelation times of a fluctuating synaptic input current \cite{Brunel2001}. Since fluctuating synaptic inputs arriving at the distal dendrite are low-pass filtered, the autocorrelation time of the corresponding input current felt by the soma is increased (or rather limited from below). 
Spike rate resonance frequencies were lower for EIF neurons as compared to LIF neurons, in particular for background inputs only at the distal dendrite. This may be explained by the fact that the presence of the exponential term, describing the spike initiating sodium current, decreases the rate response to high frequency inputs \cite{Fourcaud2003} (see also the analytical results in \cite{Ostojic2015}). 
In all cases, the amplitude of the modulation also depended on the input strength (input mean and noise intensity), but its relationship to field frequency was not strongly affected by the input parameters. %
Recently it has been shown that Purkinje neurons, due to their large dendritic trees, exhibit spike rate resonance at rather high frequencies in response to somatic input modulations and in the presence of noisy dendritic input \cite[Fig. 5A]{Ostojic2015}, which is qualitatively similar to the field-induced resonance effects described here (cf. Figs.~{\ref{fig_fRates_dend}} and {\ref{fig9}}C).  %
It should be noted, however, that an oscillatory (spatially uniform) external field corresponds to oscillatory input currents with opposite sign at the soma and the distal dendrite, respectively (cf. Eqs.~{\ref{eq:BSmodel_BC_soma}} and {\ref{eq:BSmodel_BC_end}}). The effects of the field can thus not be easily anticipated from those of an input current modulation at the soma alone. Furthermore, the dendritic membrane surface compared to the somatic one for Purkinje cells \cite{Ostojic2015} is substantially larger than that of pyramidal neurons as considered here, which additionally impedes to directly relate the results.

Existing experimental studies on the modulation of neuronal activity by extracellular fields have considered a small number of field frequencies (see \cite{Reato2013} for a review). Therefore, our results on spike rate resonance are currently not completely confirmed and may be regarded as predictions.
In accordance with our findings weak alternating electric fields (of 30~Hz) have been shown to increase the spiking coherence of pyramidal cells in rat hippocampal slices \cite{Radman2007}, where this increase was proportional to the subthreshold membrane polarization. %
Moreover, spatially uniform extracellular fields with high-frequency components entrained spiking activity in ferret primary visual cortex more effectively than fields that only contain low-frequency components \cite[Fig. S6]{Frohlich2010}.
Our predictions on spike rate modulation by an oscillating electric field are thus in agreement with current knowledge and are informative for future experimental studies. Those results may further be of potential interest for the design of transcranial electrical stimulation protocols.
Regarding the point model extension, we analytically derived an expression for an input current to reproduce the effect of the field as extracted from the biophysically grounded BS model. The amplitude and phase of this %
input current depend on the parameters of the BS neuron and the electric field. 
Previously, simple phenomenologically obtained input currents have been used for point neuron network simulations, with either constant amplitudes (across frequencies) \cite{Frohlich2010,Ali2013} or amplitudes fitted to electrophysiological data \cite{Reato2010}.
Interestingly, the latter study used an input current whose magnitude decreases with increasing frequency, in contrast to the equivalent current we obtained (whose magnitude increases with frequency up to 10~kHz). The neuronal subthreshold sensitivity in that study and the ones shown here, however, are similar. This apparent discrepancy in the currents describing the field effect may be explained by the impedance of the applied model neurons, which naturally influences the equivalent input current.
In \cite{Reato2010} the model parameters (and thus the impedance) were not fitted to real cells; hence it is unlikely that the model impedance matched with the impedance of the cells from which the current amplitudes were estimated \cite{Deans2007}.
The successful reproduction of the BS spike rate modulation due to the field by the eP model presented here supports the high-pass properties of the equivalent input current.

In the present study, we derived an extension for point neuron models of the LIF and EIF types. Additional model variables with slow dynamics \cite{Brette2005} may also be included in this framework, in order to reflect, for example, effects of slowly deactivating potassium channels that mediate spike rate adaptation and associated characteristic neuronal response properties \cite{Ladenbauer2012,Ladenbauer2014}. In that case, a separation of timescales argument could be used to derive the model extension.

The results we extracted from a canonical spatial neuron model provide insight into the effects of cellular morphology on synaptic input integration and the impact of extracellular electric fields on neuronal activity. In particular, the presented point model extension, which is straightforward to implement and efficient to simulate, shall give rise to comprehensive computational investigations of neuronal population activity entrainment due to transcranial stimulation. %

%
%
%
%
%
%
%
%
%
%

%
%
%
%
%
%
%
%
%
%
%
%
%
%
%

%
%
%
%
\section*{Methods}\label{sec:methods}

\subsection*{The ball-and-stick (BS) neuron model}\label{sec:methods_BS}

\paragraph*{Model derivation}
The BS neuron model consists of a finite passive dendritic cable of length $ L $ with a lumped somatic compartment at one extremity $ x=0 $ and a sealed-end at the other. We consider this neuron model exposed to synaptic inputs at the soma $ I_{\mathrm{s}}(t) $ and the distal dendritic end $ I_{\mathrm{d}}(t) $ and to an electric field $ E(t) $ (see Fig.~\ref{fig1}). The electric field is spatially uniform at the scale of the neuron, which is a valid assumption for fields induced by transcranial brain stimulation \cite{Bikson2012a}.
Assuming a homogeneous, purely ohmic medium (see \cite{Bedard2013} for the cable equation in a non-ohmic medium),
the subthreshold dynamics of the membrane voltage along the dendritic cable are governed by \cite{Rattay1986}
\begin{align}\label{eq:cableEquation}
c_{\mathrm{m}} \frac{\partial V_{\mathrm{BS}}}{\partial t} - g_{\mathrm{i}} \frac{\partial ^2 V_{\mathrm{BS}}}{\partial x ^ 2} + g_{\mathrm{m}} V_{\mathrm{BS}} &= - g_{\mathrm{i}} \frac{\partial E}{\partial x} = 0 \qquad 0<x<L, \\
E(t) = -\frac{\partial V_{\mathrm{BS,e}}}{\partial x}(x,t),
\end{align}
where  $ V_{\mathrm{BS}}(x,t) := V_{\mathrm{BS,i}}(x,t)-V_{\mathrm{BS,e}}(x,t)-V_{\mathrm{rest}} $, with intra- and extracellular potentials $ V_{\mathrm{BS,i}} $ and $ V_{\mathrm{BS,e}} $, respectively. $ c_{\mathrm{m}} = c D_{\mathrm{d}} \pi $ is the membrane capacitance per unit length, 
$ g_{\mathrm{i}} = \varrho_{\mathrm{i}} (D_{\mathrm{d}}/2)^2 \pi $ is the internal (axial) conductance per unit length and $ g_{\mathrm{m}} = \varrho_{\mathrm{m}}  D_{\mathrm{d}} \pi $ is the membrane conductance per unit length. $ c $ is the specific membrane capacitance (in $ \mathrm{F}/\mathrm{m}^2 $), $ \varrho_{\mathrm{i}} $ is the specific internal conductance (in $ \mathrm{S}/\mathrm{m} $), $ \varrho_{\mathrm{m}} $ is the specific membrane conductance (in $ \mathrm{S}/\mathrm{m}^2 $) and $ D_{\mathrm{d}} $ is the cable diameter.  
Note, that the rightmost equality in Eq.~\ref{eq:cableEquation} is due to our assumption of a spatially uniform electric field $ E(x,t) \equiv E(t)$. 

At the proximal end of the dendritic cable, $ x = 0 $, we consider a lumped soma, assuming that the somatic diameter $ D_{\mathrm{s}} $ is small compared to the cable length $ L $.
The corresponding boundary condition is given by \cite{Tuckwell1988}
\begin{equation}
C_{\mathrm{s}} \frac{\partial V_{\mathrm{BS}}}{\partial t} 
- g_{\mathrm{i}}\frac{\partial V_{\mathrm{BS,i}}}{\partial x} + 
G_{\mathrm{s}} V_{\mathrm{BS}} - G_{\mathrm{s}} \Delta_{\mathrm{T}} 
e^{\tfrac{V_{\mathrm{BS}}-V_{\mathrm{T}}}{\Delta_{\mathrm{T}}}} = 
I_{\mathrm{s}}(t) \qquad x=0,
\end{equation}
and thus %
\begin{equation} \label{eq:BC_soma}
C_{\mathrm{s}} \frac{\partial V_{\mathrm{BS}}}{\partial t} 
- g_{\mathrm{i}}\frac{\partial V_{\mathrm{BS}}}{\partial x} + 
G_{\mathrm{s}} V_{\mathrm{BS}} - G_{\mathrm{s}} \Delta_{\mathrm{T}} 
e^{\tfrac{V_{\mathrm{BS}}-V_{\mathrm{T}}}{\Delta_{\mathrm{T}}}} = 
I_{\mathrm{s}}(t) - g_{\mathrm{i}} E(t) \qquad x=0,
\end{equation}
where $ C_{\mathrm{s}} = c D_{\mathrm{s}}^2 \pi $ and $ G_{\mathrm{s}} = \varrho_{\mathrm{m}} D_{\mathrm{s}}^2 \pi $ are the somatic membrane capacitance and leak conductance, respectively.
At the distal end of the dendritic cable, $ x=L $, we have \cite{Tuckwell1988}
\begin{equation}
\frac{\partial V_{\mathrm{BS,i}}}{\partial x} = \frac{I_{\mathrm{d}}(t)}{g_{\mathrm{i}}} \qquad x=L,
\end{equation}
due to the synaptic input $ I_{\mathrm{d}}(t) $, and therefore,
\begin{equation} \label{eq:BC_sealed_end}
\frac{\partial V_{\mathrm{BS}}}{\partial x} = %
\frac{I_{\mathrm{d}}(t)}{g_{\mathrm{i}}} + E(t) \qquad x=L.
\end{equation}
The subthreshold voltage dynamics of the BS model are thus determined by Eqs.~\ref{eq:cableEquation}, \ref{eq:BC_soma} and \ref{eq:BC_sealed_end}.  
The spiking mechanism is implemented by the reset condition~\ref{eq:BSmodel_RC} with refractory period (see Models in the section Results).

\paragraph*{Calculation of the subthreshold somatic response}
To analytically calculate the somatic membrane voltage response of the BS model, we consider small variations of the synaptic inputs $ I_{\mathrm{s}}(t) $, $ I_{\mathrm{d}}(t) $ and a weak oscillatory electric field $ E(t) $.
This allows us to linearize the exponential term in Eq.~\ref{eq:BC_soma} around a baseline voltage value $ V_0 $ to obtain
\begin{equation} \label{eq:BC_soma_linearized}
C_{\mathrm{s}} \frac{\partial V_{\mathrm{BS}}}{\partial t} 
- g_{\mathrm{i}}\frac{\partial V_{\mathrm{BS}}}{\partial x} + 
G_{\mathrm{s}} \Big(1-e^{\tfrac{V_0-V_{\mathrm{T}}}{\Delta_{\mathrm{T}}}} \Big)
V_{\mathrm{BS}} = 
G_{\mathrm{s}}e^{\tfrac{V_0-V_{\mathrm{T}}}{\Delta_{\mathrm{T}}}}(\Delta_{\mathrm{T}} - V_0) +
I_{\mathrm{s}}(t) - g_{\mathrm{i}} E(t)
\end{equation} 
for $ x=0 $.
Note that in case of a purely leaky and capacitive neuronal membrane (i.e., without the exponential term, in the limit $ \Delta_{\mathrm{T}} \to 0 $) the linearization above is not required and the response calculated below is also exact for larger (subthreshold) synaptic inputs and electric field magnitudes.
The linear partial differential equation~\ref{eq:cableEquation} together with the boundary conditions~\ref{eq:BC_soma_linearized} and~\ref{eq:BC_sealed_end} can be solved using separation of variables $ V_{\mathrm{BS}}(x,t) = W(x) U(t) $ and the temporal Fourier transform 
\begin{equation}
\hat{V}_{\mathrm{BS}}(x,\omega) = W(x) \hat{U}(\omega) = W(x) \int_{-\infty}^{\infty} U(t) e^{\mathrm{i} \omega t} dt,
\end{equation}
where $ \omega = 2\pi f $ denotes angular frequency. We obtain the system of differential equations
\begin{align}
c_{\mathrm{m}} \mathrm{i} \omega \hat{V}_{\mathrm{BS}} - g_{\mathrm{i}} \frac{\partial ^2 \hat{V}_{\mathrm{BS}}}{\partial x ^ 2} + g_{\mathrm{m}} \hat{V}_{\mathrm{BS}} &= 0 \qquad 0<x<L, \label{eq:cableEquation_Fourier} \\
C_{\mathrm{s}} \mathrm{i} \omega \hat{V}_{\mathrm{BS}} 
- g_{\mathrm{i}}\frac{\partial \hat{V}_{\mathrm{BS}}}{\partial x} + 
G_{\mathrm{s}} \Big(1-e^{\tfrac{V_0-V_{\mathrm{T}}}{\Delta_{\mathrm{T}}}} \Big)
\hat{V}_{\mathrm{BS}} &= %
\nonumber \\
2 \pi \delta (\omega) G_{\mathrm{s}}e^{\tfrac{V_0-V_{\mathrm{T}}}{\Delta_{\mathrm{T}}}}(\Delta_{\mathrm{T}} - V_0) +
\hat{I}_{\mathrm{s}}(\omega) &- g_{\mathrm{i}} \hat{E}(\omega) \qquad x=0, \label{eq:BC_soma_Fourier} \\
\frac{\partial \hat{V}_{\mathrm{BS}}}{\partial x} &= \frac{\hat{I}_{\mathrm{d}}(\omega)}{g_{\mathrm{i}}} + \hat{E}(\omega) \qquad x=L, \label{eq:BC_sealed_end_Fourier}
\end{align}
where $ \hat{.} $ indicates the (temporal) Fourier transform and $\delta(.)$ the Dirac delta function. 
The solution of the second order linear differential equation~\ref{eq:cableEquation_Fourier} is given by 
\begin{equation} \label{eq:V_BS_Fourier}
\hat{V}_{\mathrm{BS}}(x,\omega) = a_1 (\omega) e^{z(\omega)x} + a_2 (\omega) e^{-z(\omega)x},
\end{equation}
where $ \pm z(\omega) $ are the roots of the characteristic polynomial $ g_{\mathrm{i}} \lambda^2 = g_{\mathrm{m}} + c_{\mathrm{m}} \mathrm{i} \omega $ of Eq.~\ref{eq:cableEquation_Fourier}:
\begin{equation}
z(\omega) = \sqrt{\frac{g_{\mathrm{m}} + \sqrt{g_{\mathrm{m}}^2 + \omega^2 c_{\mathrm{m}}^2}}{2g_{\mathrm{i}}}} +  \mathrm{sgn}(\omega) \mathrm{i} \sqrt{\frac{-g_{\mathrm{m}} + \sqrt{g_{\mathrm{m}}^2 + \omega^2 c_{\mathrm{m}}^2}}{2g_{\mathrm{i}}}},
\end{equation}
(same as Eq.~\ref{eq:z}). %
The coefficients $ a_1 (\omega) $ and $ a_2 (\omega) $ are calculated by inserting $\hat{V}_{\mathrm{BS}}(x,\omega)$ from Eq.~\ref{eq:V_BS_Fourier} in Eqs.~\ref{eq:BC_soma_Fourier} and \ref{eq:BC_sealed_end_Fourier} to finally obtain
\begin{equation}
\hat{V}_{\mathrm{BS}}(0,\omega) = \hat{V}_{\mathrm{BS}}^{I_{\mathrm{s}}}(0,\omega) + \hat{V}_{\mathrm{BS}}^{I_{\mathrm{d}}}(0,\omega) + \hat{V}_{\mathrm{BS}}^{\Delta_{\mathrm{T}}}(0,\omega) + \hat{V}_{\mathrm{BS}}^E(0,\omega)
\end{equation}
with
\begin{align}
\hat{V}_{\mathrm{BS}}^{I_{\mathrm{s}}}(0,\omega) &= \frac{\hat{I}_{\mathrm{s}}(\omega)}{X(\omega)}, &
\hat{V}_{\mathrm{BS}}^{\Delta_{\mathrm{T}}}(0,\omega) &= \frac{2\pi \delta(\omega) G_{\mathrm{s}}e^{\tfrac{V_0-V_{\mathrm{T}}}{\Delta_{\mathrm{T}}}}(\Delta_{\mathrm{T}} - V_0)}{X(\omega)}, \label{eq:V_BS_Is} \\
\hat{V}_{\mathrm{BS}}^{I_{\mathrm{d}}}(0,\omega) &= \frac{\hat{I}_{\mathrm{d}}(\omega) \, \mathrm{sech}(z(\omega) L)}{X(\omega)}, &
\hat{V}_{\mathrm{BS}}^E(0,\omega) &= \frac{\hat{E}(\omega) g_{\mathrm{i}} \,[\mathrm{sech}(z(\omega) L) -1]}{X(\omega)}, \label{eq:V_BS_E}
\end{align}
and 
\begin{equation}
X(\omega) := C_{\mathrm{s}} \mathrm{i} \omega + G_{\mathrm{s}}\Big(1-e^{\tfrac{V_0-V_{\mathrm{T}}}{\Delta_{\mathrm{T}}}} \Big) + z(\omega) \, g_{\mathrm{i}} \mathrm{tanh}(z(\omega) L). \label{eq:V_BS_X}
\end{equation}
The function $ \mathrm{sech}(x) = \cosh(x)^{-1}$ refers to the hyperbolic secant. 
Here, $ V_{\mathrm{BS}}^{I_{\mathrm{s}}} $, $ V_{\mathrm{BS}}^{I_{\mathrm{d}}} $ and $ V_{\mathrm{BS}}^E $ denote respectively the voltage response components to $ I_{\mathrm{s}} $, $ I_{\mathrm{d}} $ and $ E $. $ V_{\mathrm{BS}}^{\Delta_{\mathrm{T}}} $ is the voltage ``response'' caused by the (linearized) exponential term. 
Since $ E(t) = E_1 \sin(\varphi t) $, the response to the field can be expressed in the time domain as
\begin{align}
V_{\mathrm{BS}}^E (0,t) &= | A(\varphi) | \, \sin \! \big(\varphi t + \arg(A(\varphi))\big), \\ 
A(\varphi) &= \frac{E_1 g_{\mathrm{i}} \,[\mathrm{sech}(z(\varphi) L) -1]}{ C_{\mathrm{s}} \mathrm{i} \varphi + G_{\mathrm{s}}\Big(1-e^{\tfrac{V_0-V_{\mathrm{T}}}{\Delta_{\mathrm{T}}}} \Big) + z(\varphi) \, g_{\mathrm{i}} \tanh{(z(\varphi) L)} }.
\end{align} 

\subsection*{The extended point (eP) neuron model}
The subthreshold voltage dynamics of the eP model is specified by Eq.~\ref{eq:Pmodel} which is complemented by the reset condition~\ref{eq:Pmodel_RC} together with a refractory period (see Models in the section Results).

\paragraph*{Calculation of the subthreshold response}
We consider again small variations of the synaptic inputs $ I_{\mathrm{s}}(t) $, $ I_{\mathrm{d}}(t) $ and the current $ I_E(t) $ that corresponds to a weak oscillatory electric field $ E(t) $.
Linearizing the exponential term in Eq.~\ref{eq:Pmodel} around the steady-state somatic voltage value, $ V_0 $, we obtain
\begin{equation} \label{eq:Pmodel_linearized}
	C_{\mathrm{eP}} \frac{\partial V_{\mathrm{eP}}}{\partial t} 
	+ 
	G_{\mathrm{eP}} \Big(1-\alpha e^{\tfrac{V_0-V_{\mathrm{T}}}{\Delta_{\mathrm{T}}}} \Big)
	V_{\mathrm{eP}} = 
	G_{\mathrm{eP}} \alpha e^{\tfrac{V_0-V_{\mathrm{T}}}{\Delta_{\mathrm{T}}}}(\Delta_{\mathrm{T}} - V_0) +
	[L_{\mathrm{s}} \ast I_{\mathrm{s}}](t) + [L_{\mathrm{d}} \ast I_{\mathrm{d}}](t) + I_E(t).
\end{equation} 
Note that here again in case of a purely leaky and capacitive membrane ($ \Delta_{\mathrm{T}} \to 0 $) the linearization above is not required and the response calculated below is also exact for larger (subthreshold) inputs.
Using the Fourier transform on Eq.~\ref{eq:Pmodel_linearized} yields
\begin{align}
&C_{\mathrm{eP}} \mathrm{i} \omega \hat{V}_{\mathrm{eP}} 
+ 
G_{\mathrm{eP}} \Big(1-\alpha e^{\tfrac{V_0-V_{\mathrm{T}}}{\Delta_{\mathrm{T}}}} \Big)
\hat{V}_{\mathrm{eP}} = &
\nonumber \\
&2\pi \delta(\omega) 
G_{\mathrm{eP}}\alpha e^{\tfrac{V_0-V_{\mathrm{T}}}{\Delta_{\mathrm{T}}}}(\Delta_{\mathrm{T}} - V_0) +
\hat{L}_{\mathrm{s}}(\omega) \hat{I}_{\mathrm{s}}(\omega) + \hat{L}_{\mathrm{d}}(\omega) \hat{I}_{\mathrm{d}}(\omega) + \hat{I}_E(\omega),	&
\end{align}
which can be easily solved to obtain
\begin{equation}\label{eq:V_eP}
\hat{V}_{\mathrm{eP}}(\omega) = \frac{\hat{L}_{\mathrm{s}}(\omega) \hat{I}_{\mathrm{s}}(\omega) + 
\hat{L}_{\mathrm{d}}(\omega) \hat{I}_{\mathrm{d}}(\omega) + 2\pi \delta(\omega) G_{\mathrm{eP}} \alpha e^{\tfrac{V_0-V_{\mathrm{T}}}{\Delta_{\mathrm{T}}}}(\Delta_{\mathrm{T}} - V_0) + \hat{I}_E(\omega)}{C_{\mathrm{eP}} \mathrm{i} \omega + G_{\mathrm{eP}} \Big(1- \alpha e^{\tfrac{V_0-V_{\mathrm{T}}}{\Delta_{\mathrm{T}}}} \Big) }.
\end{equation}

\begin{table}[!ht]
	\centering
	\caption{
		{\bf Description of parameters and applied values}}
	\begin{tabular}{|c|c|l|}
		\hline
		\bf Parameter                               & \bf Value (range)                                                                   & \bf Description                                \\
		\bf (Unit)                                    &                                                                                     &                                                \\ \hline
		$c$ ($\mathrm{F}/\mathrm{m}^ 2$)          & $1 \cdot 10^{-2}$ ~\cite{Migliore2005,Mainen1996}                                 & Specific membrane capacitance                  \\  \hline
		$\varrho_{\mathrm{m}}$    ($ \mathrm{S}/\mathrm{m}^2 $)                   & $1/2.8$ ~\cite{Migliore2005}                                                      & Specific membrane conductance                  \\ \hline
		$\varrho_{\mathrm{i}}$ ($ \mathrm{S}/\mathrm{m} $)                     & $1/1.5$  ~\cite{Migliore2005,Mainen1996}                                          & Specific internal (axial) conductance          \\ \hline
		$D_{\mathrm{s}}$    ($ \mathrm{m} $)                         & $\{5,10^*,15\} \cdot 10^{-6}$ \cite{Rattay1999}                                   & Soma diameter                                  \\ \hline
		$D_{\mathrm{d}}$ ($ \mathrm{m} $)                              & $\{0.6,1.2^*,1.8\} \cdot 10^{-6}$ ~\cite{Major1994}                               & Dendritic cable diameter                       \\ \hline
		$L$     ($ \mathrm{m} $)                                       & $\{3.5,7^*,10.5\} \cdot 10^{-4}$ \cite{Spruston2009} & Dendritic cable length                         \\ \hline
		$ C_{\mathrm{s}} $($ \mathrm{F} $)                          & $c D_{\mathrm{s}}^2 \pi $                                                           & Somatic membrane capacitance                   \\ \hline
		$ G_{\mathrm{s}} $    ($ \mathrm{S} $)                          & $\varrho_{\mathrm{m}} D_{\mathrm{s}}^2 \pi$                                         & Somatic membrane conductance                   \\ \hline
		$ c_{\mathrm{m}} $   ($ \mathrm{F}/\mathrm{m} $)                          & $c D_{\mathrm{d}} \pi $                                                             & Dendritic membrane capacitance \\
		             &                                                                                     &                                          per unit length       \\ \hline
		$ g_{\mathrm{m}} $    ($ \mathrm{S}/\mathrm{m} $)                       & $\varrho_{\mathrm{m}} D_{\mathrm{d}} \pi$                                           & Dendritic membrane conductance \\
		               &                                                                                     &                                        per unit length         \\ \hline
		$ g_{\mathrm{i}} $ ($ \mathrm{S}\cdot\mathrm{m} $)                            & $\varrho_{\mathrm{i}} (D_{\mathrm{d}}/2)^2 \pi$                                     & Internal (axial) conductance   \\
		           &                                                                                     &                                  per unit length               \\ \hline
		$ V_{\mathrm{s}} $     ($ \mathrm{mV} $)                       & $\{10, 20\}$                                                                        & Spike (or cutoff) voltage                      \\ \hline
		$ V_{\mathrm{r}} $        ($ \mathrm{mV} $)                      & $0$                                                                                 & Reset voltage of BS model                      \\ \hline
		$ V_{\mathrm{T}} $     ($ \mathrm{mV} $)                         & $10$ ~\cite{Badel2008JN}                                                          & Threshold voltage                              \\ \hline
		$ V_{0} $     ($ \mathrm{mV} $)                                   &$V_\mathrm{r}$                                                                              & Baseline voltage for EIF model extension  \\ \hline
		$ \Delta_{\mathrm{T}} $     ($ \mathrm{mV} $)                   & $1.5$ ~\cite{Badel2008JN}                                                         & Threshold slope factor                         \\ \hline
		$ T_{\mathrm{Ref}} $   ($ \mathrm{ms} $)                         & $1.5$                                                                               & Duration of refractory period                  \\ \hline

		$ C_{\mathrm{eP}} $ ($ \mathrm{F} $)                           & $ C_{\mathrm{s}} $                                                                  & Membrane capacitance of eP model               \\ \hline
		$ G_{\mathrm{eP}} $ 	($ \mathrm{S} $)                          & $ G_{\mathrm{s}} $                                                                  & Membrane conductance of eP model               \\ \hline
		$ V_{\mathrm{r}}^\prime $    	($ \mathrm{mV} $)                   & $ 5 $                                                                               & Reset voltage of eP and P models               \\ \hline
		$ I_\mathrm{s}^0 $    ($ \mathrm{pA} $)                                  & $[4.254, 11.407]$                                                                   & Mean input current at the soma                    \\ \hline
		$ \sigma_\mathrm{s} $       ($ \mathrm{pA} $)                              & $[8.887, 74.512]$                                                                   & Somatic input noise intensity                 \\ \hline
		$ I_\mathrm{d}^0 $    ($ \mathrm{pA} $)                                  & $[6.255, 13.214]$                                                                   & Mean input current at the dendrite                  \\ \hline
		$ \sigma_\mathrm{d} $       ($ \mathrm{pA} $)                              & $[21.875, 122.363]$                                                                   & Dendritic input noise intensity                 \\ \hline
		$ \tau $          ($ \mathrm{ms} $)                             & $ 0.5 $                                                                             & Synaptic current correlation time              \\ \hline
		$ E_1 $      ($ \mathrm{V}/\mathrm{m} $)                                & $\{1, 10\}$                                                                         & Amplitude of electric field                   \\ \hline
		$ \varphi $      ($ \mathrm{rad} $)                           & $ [0, 10^4]\cdot 2 \pi $                                                            & Angular frequency of electric field            \\ \hline
		$ \Delta $   ($ \mathrm{ms} $)                                  & $ 3 $                                                                               & Spike coincidence precision                    \\ \hline
	\end{tabular} 
	
	$ ^* $ indicates default values.
	\label{table:BS_parameters}
\end{table}

\subsection*{Numerical simulation}\label{sec:methods_num_sim}

\paragraph{Synaptic input and electric field}
To generate spike trains we considered in-vivo like noisy synaptic inputs $ I_{\mathrm{s}}(t) $, $ I_{\mathrm{d}}(t) $. Specifically, $ I_{\mathrm{x}}(t) $, $\mathrm{x} \in \lbrace \mathrm{s},\mathrm{d} \rbrace $ was described by an Ornstein-Uhlenbeck process 
\begin{equation} \label{eq:OU_input}
\frac{d I_{\mathrm{x}}}{dt} = \frac{I_\mathrm{x}^0 - I_{\mathrm{x}}}{\tau} + \sigma_{\mathrm{x}} \sqrt{\frac{2}{\tau}} \xi_\mathrm{x}(t),
\end{equation}
with mean $ I_\mathrm{x}^0 $, correlation time $ \tau $ and standard deviation $ \sigma_{\mathrm{x}} $.
$ \xi_{\mathrm{x}}(t) $ is a Gaussian white noise process, i.e. with zero mean $ \langle \xi_{\mathrm{x}}(t) \rangle = 0$ and delta autocorrelation $ \langle \xi_{\mathrm{x}}(t) \xi_{\mathrm{x}}(t+t^\prime) \rangle = \delta(t^\prime)$, where $ \langle . \rangle $ denotes the expectation operator. Eq.~\ref{eq:OU_input} was numerically solved using the method described in \cite{Gillespie1996}.

The electric field was described by 
\begin{equation} \label{eq:electric_field_meth}
E(t) = E_1 \sin (\varphi t),
\end{equation}
(same as Eq.~\ref{eq:electric_field}) with amplitude $ E_1 $ and angular frequency $ \varphi = 2 \pi f $. The values for all parameters are provided in Table~\ref{table:BS_parameters}.

\paragraph{Ball-and-stick neuron model}
The BS neuron model was numerically solved using the NEURON simulation environment \cite{Hines2001}. We applied a finite difference space discretization scheme with 50 segments for the dendritic cable and the implicit (or backward) Euler time discretization scheme. The integration time step was fixed to 0.05~ms when the exponential term was omitted ($\Delta_\mathrm{T} \to 0 $) and 0.025~ms otherwise. 
Decreasing the time step size and increasing the number of segments did not lead to noticeable changes in the membrane voltage time series.
The time-varying extracellular potential was included using the built-in ``extracellular'' mechanism in NEURON. 

\paragraph{Point neuron model}

The point neuron models were numerically solved using the forward Euler time discretization scheme and the same time step as used for the BS model. The solution method was implemented in Python using the library ``Numba'' for fast computation. %
Both point model neurons received the same realization of synaptic input $ I_{\mathrm{x}}(t) $, $\mathrm{x} \in \lbrace \mathrm{s},\mathrm{d} \rbrace $ as the BS model neuron. The linear filters $ L_{\mathrm{x}}(t) $, $\mathrm{x} \in \lbrace \mathrm{s},\mathrm{d} \rbrace $ in the eP model were implemented using the fast Fourier transform (FFT) of $ I_{\mathrm{x}}(t) $ and the inverse FFT of $ \hat{L}_{\mathrm{x}}(\omega) \hat{I}_{\mathrm{x}}(\omega) $.
The membrane capacitance and conductance of the eP model were chosen as equal to the corresponding somatic quantities of the BS model, $ C_{\mathrm{eP}}=C_{\mathrm{s}} $, $ G_{\mathrm{eP}} = G_{\mathrm{s}} $.
Except stated otherwise, the corresponding parameter values of the P model were determined by fitting the BS model spiking activity in terms of spike coincidences. Specifically, $ G_{\mathrm{P}} $ was chosen such that the steady-state (somatic) membrane voltage of the BS model is matched exactly, i.e. $ Z_{\mathrm{P}}^\mathrm{x}(0) = Z_{\mathrm{BS}}^\mathrm{x}(0) $, $\mathrm{x} \in \lbrace \mathrm{s},\mathrm{d} \rbrace $, with impedance $ Z_{\mathrm{m}}^\mathrm{x}(\omega) := \hat{V}_{\mathrm{m}}^{I_\mathrm{x}}(\omega)/\hat{I}_{\mathrm{x}}(\omega)$, $\mathrm{m} \in \{ \mathrm{BS}|_{x=0},\mathrm{P} \}$ using Eqs.~\ref{eq:BS_somatic_V_resp}, \ref{eq:BS_somatic_V_resp_Id} and \ref{eq:P_somatic_V_resp}.
The value for $ C_{\mathrm{P}} $ was then selected to maximize the coincidence factor $ \Gamma $ (defined below) between 52~s lasting spike trains of the BS and P model neurons for each shown pair ($ I_\mathrm{x}^0 $, $ \sigma_\mathrm{x} $) of input parameter values.
In presence of the exponential term in the models ($ \Delta_{\mathrm{T}} > 0$) we used $ V_0 = V_{\mathrm{r}}$, which is the steady-state (somatic) voltage in the absence of synaptic input. %
Using a different value for $ V_0 $ did not lead to a substantial improvement of the reproduction performance. Several values were tested in the range $ [V_{\mathrm{r}}, V_{\mathrm{s}}] $ (results not shown).

\subsection*{Analysis methods for the spike trains}

\paragraph*{Spike coincidence measure}\label{sec:methods_gamma}
To quantify the similarity between the spike trains of the different model neurons we used the coincidence factor $ \Gamma $ defined by \cite{Gerstner2002} 
\begin{equation}
\Gamma_{\mathrm{ref},\mathrm{comp}} = \frac{N_\mathrm{coinc} - \langle N_\mathrm{coinc} \rangle}{(N_\mathrm{ref} + N_\mathrm{comp})/2} \, \frac{1}{\mathcal{N}},
\end{equation}
where $ N_\mathrm{coinc} $ is the number of coincident spikes with precision (i.e., maximal temporal separation) $\Delta$, $ N_\mathrm{ref} $ and $ N_\mathrm{comp} $ are the number of spikes in the reference spike train and in the one being compared to it, respectively. $\langle N_\mathrm{coinc} \rangle = 2 r \Delta N_\mathrm{ref}$ is the expected number of coincidences generated by a homogeneous Poisson process with the same spike rate $ r = N_\mathrm{comp}/T $ as that shown by the compared spike train, where $ T $ is the spike train duration.
The factor $ \mathcal{N} = 1-2 r \Delta$ normalizes $ \Gamma_{\mathrm{ref},\mathrm{comp}} $ to a maximum value of 1 which is reached if the spike trains match optimally (with precision $\Delta$). $ \Gamma_{\mathrm{ref},\mathrm{comp}} = 0$ on the other hand would result from a homogeneous Poisson process with the same rate as for the reference spike train and thus indicates pure chance.

\paragraph*{Spike rate resonance and phase shift measure}\label{sec:Methods_rate_mod}
To examine and compare their suprathreshold responses to an oscillatory electric field $ E(t) $ (Eq.~\ref{eq:electric_field_meth}), we simulated the neuron models subject to both a field and a noisy synaptic background current.
The latter was located either at the soma or at the distal dendrite $ I_{\mathrm{x}}(t) $, $\mathrm{x} \in \lbrace \mathrm{s},\mathrm{d} \rbrace $ (Eq.~\ref{eq:OU_input}). We considered regimes (in terms of ($ I_\mathrm{x}^0 $, $ \sigma_\mathrm{x} $)-pairs) where the synaptic drive is sufficiently strong to cause the neuron to spike stochastically with rate $ r_0 $. 
The sinusoidal field then causes a modulation of the spike rate that becomes apparent over many trials (i.e., independent realizations of $ I_{\mathrm{x}}(t) $). This quantity can also be thought of as the spike rate averaged over a population of neurons which individually receive a noisy synaptic drive but collectively respond to the same oscillatory field. This population, or trial-averaged, instantaneous spike rate can be expressed as
\begin{equation}
r(t) = r_0 + r_1(\varphi) \sin(\varphi t + \psi(\varphi)),
\end{equation}
where $ r_1 $ and $ \psi $ denote respectively the amplitude and phase shift, both depending on the angular frequency $ \varphi $ of the field.
Note that in the eP model the field effect is described by the oscillatory current $ I_E(t) $.
To estimate the spike rate modulation at a given field frequency we first extracted and collected the field phase $ \phi_\mathrm{s} \in [0, 2\pi)$ for each spike time $ t_\mathrm{s} $, such that $ E(t_\mathrm{s}) = E_0 \sin(\phi_\mathrm{s}) $. 
These phases were calculated from 944 trials of at least 26~s duration each, for which the first 2~s were disregarded to avoid transients, and only complete field cycles were considered.
We then computed a spike rate histogram from the set $ \{ \phi_\mathrm{s} \} $ using 20 equally sized bins and 
finally applied a sinusoid of the form $ F(\phi) = r_0 + r_1 \sin(\phi + \psi) $ with $ \phi \in [0, 2\pi) $ 
to fit that histogram using the method of least squares, where $ r_0 $ was given by the histogram mean value. In this way we obtained $ r_1 $ and $ \psi $.
%

%
%
%
%
%
%
%
%
%
%

\section*{Acknowledgments}

This work was supported by the
Deutsche Forschungsgemeinschaft (DFG)
Collaborative Research Center SFB910 (JL, KO)
and the DFG Priority Programme SPP1665 (FA,
KO). The funders had no role in study design, data collection and analysis, decision to publish, or preparation of the manuscript.

\newpage
\section*{Supplementary Figures}\label{sec:supps}

\renewcommand{\figurename}{Supplementary Figure}
\setcounter{figure}{0}

\begin{figure}[!ht]
	\begin{center}
		\includegraphics[width=0.75\textwidth]{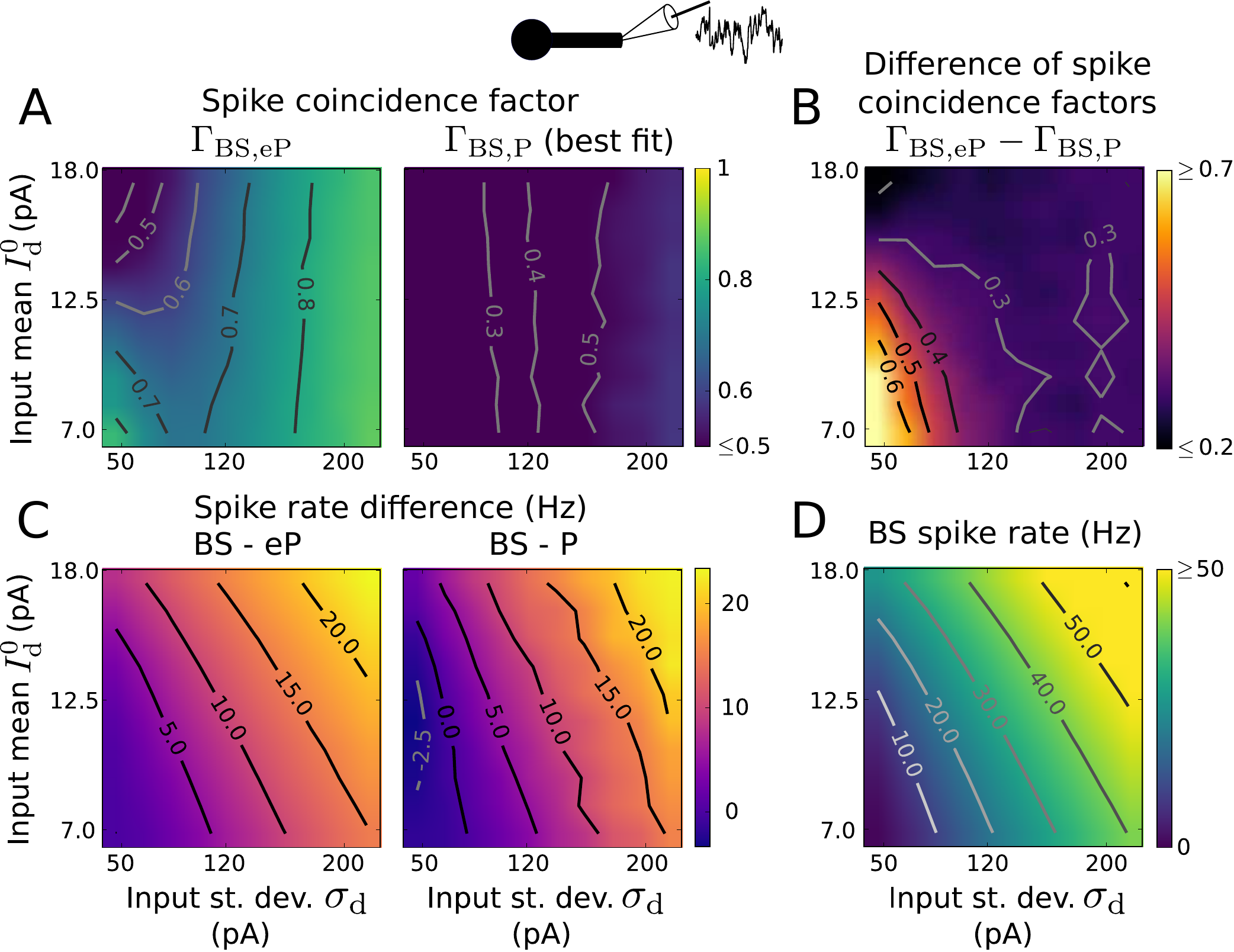}
	\end{center}
	\caption{{\bf Reproduction of spiking activity for dendritic inputs using EIF type models}
		A: Coincidence factor for the BS and eP model spike trains, $\Gamma_{\mathrm{BS, eP}}$ (left), and for the BS and P model spike trains, $\Gamma_{\mathrm{BS, P}}$ (left) as a function of input mean $ I_\mathrm{d}^0 $ and standard deviation $ \sigma_\mathrm{d} $. The parameter values of the P model were optimized to maximize $\Gamma_{\mathrm{BS, P}}$ for each input (i.e., ($ I_\mathrm{d}^0 $, $ \sigma_\mathrm{d} $)-pair) independently. 
		B: Difference $\Gamma_{\mathrm{BS, eP}} - \Gamma_{\mathrm{BS, P}}$ between the coincidence factors shown in B.
		C: Spike rate difference of the BS and eP models (left) and of the BS and P models (right) as a function of $ I_\mathrm{d}^0 $ and $ \sigma_\mathrm{d} $. 
		D: Spike rate of the BS neuron model.
		Results presented in A-D show averages over 6 noise realizations. The parameter values of the BS model are listed in Table~1.
	}
	\label{Supp_info_gamma_EIF_overfitting}
\end{figure}

\begin{figure}[!ht]
	\begin{center}
		\includegraphics[width=0.9\textwidth]{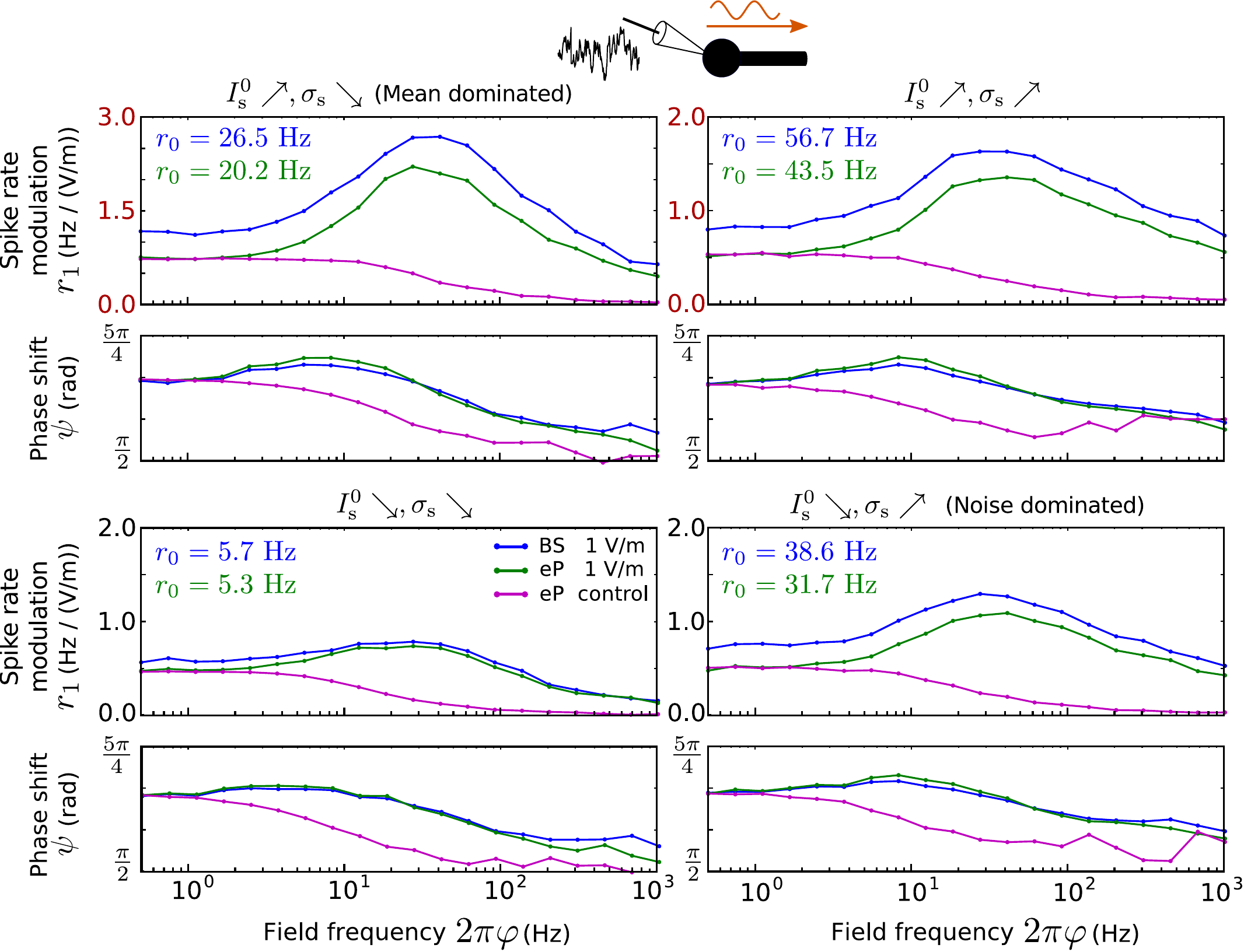}
	\end{center}
	\caption{{\bf Spike rate modulation due to an electric field for somatic inputs using neuron models of the EIF type}
		Spike rate modulation of the BS (blue) and the eP (green) models due to an oscillating electric field ($ E_1=1$~V/m) as a function of its frequency, for different somatic inputs: $ I_\mathrm{s}^0 = 10.61 $~pA,  $\sigma_\mathrm{s} = 24.08 $~pA (top left), $ I_\mathrm{s}^0 = 10.61 $~pA, $\sigma_\mathrm{s} = 68.21 $~pA (top right), $ I_\mathrm{s}^0 = 5.05 $~pA, $\sigma_\mathrm{s} = 24.08$~pA (bottom left), and $ I_\mathrm{s}^0 = 5.05 $~pA, $\sigma_\mathrm{s} = 68.21 $~pA (bottom right).
		Magenta lines show the spike rate modulation of the eP model for which $ I_E $ was given by $ I_E(t) = I_1 \sin(\varphi t + \phi) $ with constant amplitude $ I_1 = |B(0.5/(2\pi))|$ and phase shift $\phi = \arg\left(B(0.5/(2\pi))\right)$ with $B$ from Eq.~21 and $ E_1=10 $~V/m.
		Note the different amplitude scales in the two top plots.
	}	
	\label{Supp_info_rate_mod_field_EIF}
\end{figure}

\begin{figure}[!ht]
	\begin{center}
		\includegraphics[width=0.9\textwidth]{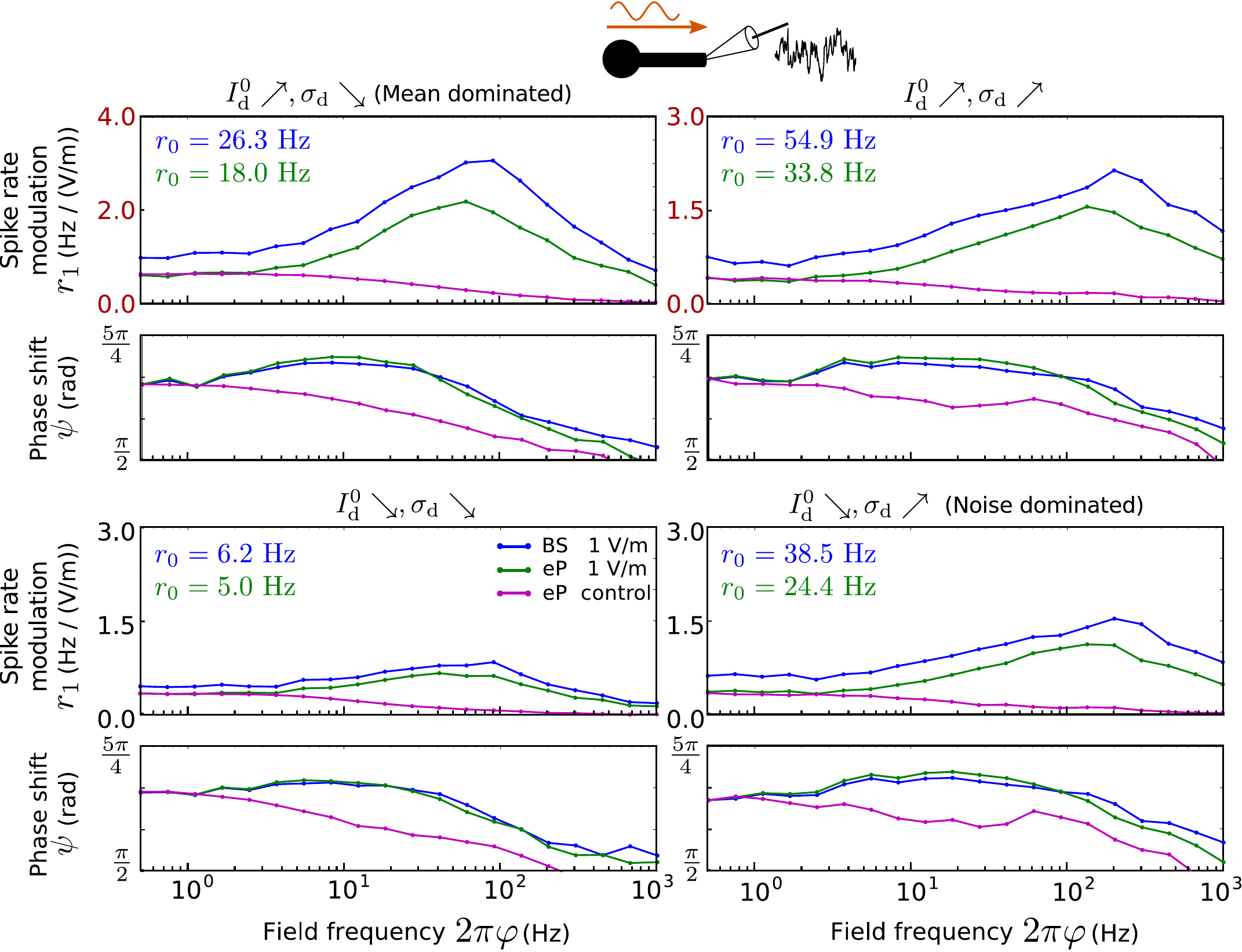}
	\end{center}
	\caption{{\bf Spike rate modulation due to an electric field for distal dendritic inputs using neuron models of the EIF type}
		Spike rate modulation of the BS (blue) and the eP (green) models due to an oscillating electric field ($ E_1=1$~V/m) as a function of its frequency, for different distal dendritic inputs: $ I_\mathrm{d}^0 =16.73 $~pA,  $\sigma_\mathrm{d} =57.73 $~pA (top left), $ I_\mathrm{d}^0 =16.73 $~pA, $\sigma_\mathrm{d} =203.41 $~pA (top right), $ I_\mathrm{d}^0 = 7.56 $~pA, $\sigma_\mathrm{d} = 57.73 $~pA (bottom left), and $ I_\mathrm{d}^0 =7.56 $~pA, $\sigma_\mathrm{d} =203.41 $~pA (bottom right).
		Magenta lines show the spike rate modulation of the eP model for which $ I_E $ was given by $ I_E(t) = I_1 \sin(\varphi t + \phi) $ with constant amplitude $ I_1 = |B(0.5/(2\pi))|$ and phase shift $\phi = \arg\left(B(0.5/(2\pi))\right)$ with $B$ from Eq.~21 and $ E_1=10 $~V/m.
		Note the different amplitude scales in the two top plots.
	}	
\end{figure}

\end{document}